\documentclass[pdflatex,sn-basic]{sn-jnl}

\usepackage{graphicx}
\usepackage{tikz}
\usepackage{multirow}
\usepackage{pgfplots}
\usepackage{longtable}
\usepackage{enumitem}
\usepackage{booktabs}
\usepackage{tabularx}
\usepackage{ragged2e}
\usepackage{pdflscape}
\usepackage{array}
\usepackage[table]{xcolor}
\usepackage[most]{tcolorbox}
\usepackage{ltablex}
\usepackage{subcaption}
\usepackage[rightmargin=0.3pt,leftmargin=0.3pt]{quoting}
\usepackage{listings} 
\usepackage{makecell}

\keepXColumns
\usepackage{subcaption}
\usepackage{listings}
\usepackage[rightmargin=0.3pt,leftmargin=0.3pt]{quoting}

\theoremstyle{thmstyleone}%

\theoremstyle{thmstyletwo}%

\theoremstyle{thmstylethree}%

\newcolumntype{Y}{>{\RaggedRight\arraybackslash}X}
\newcommand{\SCQ}{SCQ}     
\newcommand{\MCQ}{MCQ}     
\newcommand{\LIK}{Likert}  
\newcommand{\FT}{Free-text}

\definecolor{Extremely}{HTML}{D55E00}   
\definecolor{Very}{HTML}{F4A261}        
\definecolor{Moderately}{HTML}{E6E6E6}  
\definecolor{Slightly}{HTML}{56B4E9}    
\definecolor{NotAtAll}{HTML}{0072B2}    

\definecolor{PBNew}{HTML}{0072B2}       
\definecolor{PBMature}{HTML}{D55E00}    
\definecolor{PBNeutral}{HTML}{E6E6E6}   

\definecolor{FSmall}{HTML}{F4A261}      
\definecolor{FMedium}{HTML}{D55E00}     
\definecolor{FLarge}{HTML}{56B4E9}      
\definecolor{FNeutral}{HTML}{E6E6E6}    

\newtcolorbox{surveybox}{
  enhanced,
  arc=1mm,
  colback=white,
  colframe=black,
  boxrule=0.7pt,
  width=\textwidth, 
  left=4mm,
  right=4mm,
  top=3mm,
  bottom=3mm,
  before skip=10pt,
  after skip=10pt,
}

\pgfplotsset{
  likertGrey/.style={
    xbar stacked,
    xmin=0, xmax=100,
    xtick={0,20,40,60,80,100},
    axis x line*=bottom, 
    axis y line*=left,
    xtick pos=lower,
    ytick pos=left,
    tick align=outside,
    xticklabel style={yshift=-4pt},        
    enlarge y limits=true,
    clip=false, 
    y=0.75cm,
    bar width=0.32cm,
    tick label style={font=\small},
    label style={font=\small},
    title style={font=\small, yshift=-2pt},

    cycle list={
  {color=NotAtAll,   fill=NotAtAll,   draw=black!60},
  {color=Slightly,   fill=Slightly,   draw=black!60},
  {color=Moderately, fill=Moderately, draw=black!60},
  {color=Very,       fill=Very,       draw=black!60},
  {color=Extremely,  fill=Extremely,  draw=black!60},
},
    axis line style={black!50},
    tick style={black!50},
  }
}

\pgfplotsset{
  likertLegendBoxes/.style={
    legend image code/.code={
      \draw[#1, draw=black!35] (0cm,-0.08cm) rectangle (0.28cm,0.18cm);
    }
  }
}

\newcommand{\progressbar}[7]{%
\begin{tikzpicture}[baseline=(current bounding box.center)]
  \def\H{0.23cm}

  \def\L{1.4cm}
  \def\R{1.4cm}

  \def\G{0.25cm}

  \pgfmathsetlengthmacro{\W}{\linewidth-\L-\R-2*\G}
  \pgfmathsetlengthmacro{\Y}{0.5*\H}

  \node[anchor=west, font=\small] at (0,\Y) {\makebox[\L][r]{#6}};

  \pgfmathsetlengthmacro{\X}{\L+\G}

  \fill[Extremely]   (\X,0) rectangle ({\X+\W*(#1/100)},\H);
  \fill[Very]        ({\X+\W*(#1/100)},0) rectangle ({\X+\W*((#1+#2)/100)},\H);
  \fill[Moderately]  ({\X+\W*((#1+#2)/100)},0) rectangle ({\X+\W*((#1+#2+#3)/100)},\H);
  \fill[Slightly]    ({\X+\W*((#1+#2+#3)/100)},0) rectangle ({\X+\W*((#1+#2+#3+#4)/100)},\H);
  \fill[NotAtAll]    ({\X+\W*((#1+#2+#3+#4)/100)},0) rectangle ({\X+\W},\H);

  \draw[black!40] (\X,0) rectangle ({\X+\W},\H);

  \node[anchor=west, font=\small] at ({\X+\W+\G},\Y) {\makebox[\R][l]{#7}};
\end{tikzpicture}%
}

\newcommand{\progressbarThree}[5]{%
\begin{tikzpicture}[baseline=(current bounding box.center)]
  \def\H{0.23cm}

  \def\L{1.4cm}
  \def\R{1.4cm}
  \def\G{0.25cm}

  \pgfmathsetlengthmacro{\W}{\linewidth-\L-\R-2*\G}
  \pgfmathsetlengthmacro{\Y}{0.5*\H}

  \node[anchor=west, font=\small] at (0,\Y) {\makebox[\L][r]{#4}};

  \pgfmathsetlengthmacro{\X}{\L+\G}

\fill[PBNew]     (\X,0) rectangle ({\X+\W*(#1/100)},\H); 
\fill[PBMature]  ({\X+\W*(#1/100)},0) rectangle ({\X+\W*((#1+#2)/100)},\H); 
\fill[PBNeutral] ({\X+\W*((#1+#2)/100)},0) rectangle ({\X+\W},\H); 

  \draw[black!40] (\X,0) rectangle ({\X+\W},\H);

  \node[anchor=west, font=\small] at ({\X+\W+\G},\Y) {\makebox[\R][l]{#5}};
\end{tikzpicture}%
}

\newcommand{\progressbarFour}[6]{%
\begin{tikzpicture}[baseline=(current bounding box.center)]
  \def\H{0.23cm}

  \def\L{1.4cm}
  \def\R{1.4cm}
  \def\G{0.25cm}

  \pgfmathsetlengthmacro{\W}{\linewidth-\L-\R-2*\G}
  \pgfmathsetlengthmacro{\Y}{0.5*\H}

  \node[anchor=west, font=\small] at (0,\Y) {\makebox[\L][r]{#5}};

  \pgfmathsetlengthmacro{\X}{\L+\G}

\fill[FSmall]   (\X,0) rectangle ({\X+\W*(#1/100)},\H); 
\fill[FMedium]  ({\X+\W*(#1/100)},0) rectangle ({\X+\W*((#1+#2)/100)},\H); 
\fill[FLarge]   ({\X+\W*((#1+#2)/100)},0) rectangle ({\X+\W*((#1+#2+#3)/100)},\H); 
\fill[FNeutral] ({\X+\W*((#1+#2+#3)/100)},0) rectangle ({\X+\W},\H); 

  \draw[black!40] (\X,0) rectangle ({\X+\W},\H);

  \node[anchor=west, font=\small] at ({\X+\W+\G},\Y) {\makebox[\R][l]{#6}};
\end{tikzpicture}%
}

\newcommand{\legendboxA}[1]{\tikz[baseline=-0.6ex]\fill[#1] (0,0) rectangle (0.28,0.20);}

\newcommand{\legendbox}[1]{\tikz\fill[#1] (0,0) rectangle (0.25,0.18);}

\newcommand{\likertlegend}{
\begin{center}
\scriptsize
\begin{tabular}{@{}cc@{\hspace{0.6em}}cc@{\hspace{0.6em}}cc@{\hspace{0.6em}}cc@{\hspace{0.6em}}cc@{}}
\textcolor{Extremely}{\rule{0.8em}{1.8em}} & \shortstack[l]{Extremely\\important} &
\textcolor{Very}{\rule{0.8em}{1.8em}} & \shortstack[l]{Very\\important} &
\textcolor{Moderately}{\rule{0.8em}{1.8em}} & \shortstack[l]{Moderately\\important} &
\textcolor{Slightly}{\rule{0.8em}{1.8em}} & \shortstack[l]{Slightly\\important} &
\textcolor{NotAtAll}{\rule{0.8em}{1.8em}} & \shortstack[l]{Not at all\\important}
\end{tabular}
\end{center}
}

\begin{document}

\title[Article Title]{What Motivates Whom? A Survey of Newcomers to OSS and Experienced OSS Practitioners}

\author*[1]{\fnm{Shashiwadana} \sur{Nirmani}}\email{s.dona@research.deakin.edu.au}

\author[1]{\fnm{Hourieh} \sur{Khalajzadeh}}\email{hourieh.khalajzadeh@deakin.edu.au}

\author[2]{\fnm{Mojtaba} \sur{Shahin}}\email{mojtaba.shahin@rmit.edu.au}

\author[1]{\fnm{Xiao} \sur{Liu}}\email{xiao.liu@deakin.edu.au}

\affil[1]{\orgdiv{School of Information Technology}, \orgname{Deakin University}, \orgaddress{\country{Australia}}}

\affil[2]{\orgdiv{School of Computing Technologies}, \orgname{RMIT University}, \orgaddress{\country{Australia}}}

\abstract{ Open source software (OSS) development continues to expand, yet software practitioners often struggle to select suitable projects, leading to inefficient onboarding and disengagement. 
\textcolor{black}{Understanding how contributors select OSS projects is important for supporting contributors onboarding, engagement, and long-term participation within OSS communities. This study investigates contributors’ project-selection preferences in OSS projects and examines how these preferences correlate with contributors’ motivations and demographic backgrounds}. Through an online survey of 208 practitioners, we found that demographic factors, such as age, gender, and the OSS role they held, \textcolor{black}{significantly correlate with} their motivations. Additionally, preferences for project characteristics such as project age, development stage, and documentation quality vary based on specific motivations. Importantly, our findings are presented through a comparative lens, analyzing the responses of newcomers to OSS and experienced OSS practitioners separately to uncover their distinct preferences. Lastly, we explore software practitioners’ perspectives on \textcolor{black}{how existing recommendation systems could better support project selection and align with their motivations}. By disentangling the unique needs of newcomers to OSS and OSS practitioners, our findings provide insights for researchers, OSS project owners, and software practitioners to \textcolor{black}{improve contributor onboarding, engagement, and retention, while also informing future project recommendation systems} and improving the OSS ecosystem.}


\maketitle
\section{Introduction}

Open source software (OSS) development has become a widely adopted approach in the software industry \citep{yue2022off}. It engages contributors from all over the world, forming a vibrant, distributed community \citep{trinkenreich2024investigating, steinmacher2019overcoming}. Moreover, OSS development is shifting towards a hybrid model with co-existence of volunteers and paid contributors \citep{guizani2022attracting}. In 2024, OSS contributors worldwide made nearly 1 billion contributions \citep{github_octoverse_2024}. Despite this, software practitioners often face technical and social challenges when onboarding to OSS projects \citep{santos2024software}. Unsuccessful onboarding affects both contributors and projects, increasing the time and effort required to find suitable projects, ultimately leading contributors to disengage \citep{phatangare2024codecompass,santos2024software}. \textcolor{black}{Understanding how practitioners select OSS projects and what factors influence these decisions is therefore important for improving contributor onboarding and long-term retention in OSS communities.}

\textcolor{black}{Contributors do not select OSS projects solely based on technical characteristics. Their decisions are also influenced by human factors, such as motivations, personality traits, sentiments \citep{nirmani2024systematic}.  Developing a deeper understanding of these human factors may help OSS communities better support contributors during project selection, onboarding, and long-term engagement. Furthermore, such insights may inform future tools and approaches, including OSS project recommendation systems \citep{nirmani2024systematic}.}

Motivation, as one of the human factors, is highlighted as having the most significant impact on practitioner productivity and software quality management within the realm of Software Engineering (SE) \citep{vizcaino2025understanding}. Despite its recognized importance among both practitioners and researchers, motivation remains a persistent challenge to manage effectively \citep{yousef2024software}.  
Understanding what motivates individuals to contribute to OSS is crucial for sustaining and growing the OSS ecosystem.

Various factors can influence the motivations of OSS contributors. Prior research has shown that coders and non-coders have different career aspirations in OSS \citep{trinkenreich2020hidden}. Moreover, studies have discovered that gender and identity can discourage marginalized groups from contributing to OSS due to discrimination \citep{sultana2024assessing}. OSS supply, demand, and networking relationships vary based on the countries \citep{wright2023open}. While it is evident that demographics influence different motivations to contribute to OSS, there is a significant gap in research exploring the demographic influence on motivations. 

In addition, project characteristics such as popularity, documentation quality, and structured contribution guidelines affect contributors’ engagement. Projects with more stars, recent commits, detailed README files, and clear contribution templates are more likely to attract new contributors \citep{qiu2019signals}. Research on OSS discussion threads found that conversations centered around values such as respectfulness can occasionally motivate individuals to contribute, but they may also lead to disruptions and decreased motivation \citep{jamieson2024predicting}. These findings suggest a relationship between motivations and project characteristics, highlighting the need for further exploration of how motivations 
\textcolor{black}{correlated with} contributors' preferences for project characteristics.  

In this paper, we investigate how demographic factors, such as age, gender, and contributors' roles in OSS projects \textcolor{black}{correlate with} their motivations (RQ1), how these motivations \textcolor{black}{correlate with} project characteristics (RQ2), and how demographic factors themselves correlate with those preferences (RQ3). Further, we explore \textcolor{black}{how OSS project recommendation systems could better support project selection and align with contributors' motivations (RQ4).} To this end, we conducted an online survey with 208 software practitioners to analyze  \textcolor{black}{contributors’ project-selection preferences, how these preferences correlate with motivations and demographic backgrounds, and contributors’ perspectives on improving OSS recommendation systems.} The survey consisted of 30 questions, including both closed-ended and open-ended questions.  We employed statistical analysis and thematic analysis to analyze the survey responses. We divided respondents into two groups for analysis. ``\texttt{OSS practitioners}'' are defined as respondents with prior experience contributing to OSS projects, and ``\texttt{newcomers to OSS}'' are defined as respondents with a background in software engineering but no previous experience contributing to OSS projects. 

The results revealed that demographics, such as age and gender, as well as the role these contributors held in the OSS team, significantly correlate contributor motivations (RQ1); for example,  \textit{career advancement} motivation was rated highest among \texttt{OSS practitioners} aged \textit{18–24, postgraduates, and professionals}. Motivations, in turn, \textcolor{black}{correlated with} project characteristics (RQ2)-\texttt{Newcomers to OSS} motivated by \textit{gaining a reputation} and \textit{financial incentives} demonstrated a clear preference for projects having \textit{clear contribution guidelines.} Demographics also significantly correlated with project preferences (RQ3); for instance, among \texttt{OSS practitioners}, new projects were preferred by participants from Africa compared to Europe. Finally, practitioners expressed a need for interactive, personalized recommendation systems that reflect their motivations, growth opportunities, and collaborative preferences (RQ4).

Understanding the motivational needs of \texttt{newcomers to OSS} can help attract and onboard them more effectively, while analyzing the preferences of experienced \texttt{OSS practitioners} provides deeper insight into what sustains long-term engagement. This dual perspective enables the design of recommendation systems that not only foster initial interest but also support ongoing participation. \textcolor{black}{We expect these results to provide a deeper understanding of how motivations, demographics, and project characteristics correlate with OSS project-selection decisions.} Furthermore, this understanding can guide OSS communities in developing tools and methodologies that enhance contributor engagement and retention. Such insights can also support organizations in adopting more effective strategies to attract and sustain diverse OSS contributors, ultimately strengthening the OSS ecosystem. The main contributions of this work include:
\begin{itemize}
    \item Empirical analysis demonstrating how demographic factors \textcolor{black}{correlate with} the motivational preferences of different contributor groups.
    \item A comparative analysis of \texttt{newcomers to OSS} and experienced \texttt{OSS practitioners}, \textcolor{black}{examining how contributors with different motivations reported different project characteristic preferences during project selection.}
    \item An exploration of correlations between demographic traits and project characteristic preferences among \texttt{newcomers to OSS} and experienced \texttt{OSS practitioners}.
    \item Providing implications for OSS project owners, OSS recommendation system builders, OSS contributors, and researchers.    
\end{itemize}

The remainder of this paper is organized as follows. Section \ref{lit_review} outlines the Related Work, Section \ref{method} presents the Methodology, Section \ref{results} presents the Results, Section \ref{discussion} outlines the Discussion, and Section \ref{threats} discusses the Threats to Validity. Finally, Section \ref{conclusion} concludes the paper and points out some future work.

\section{Related Work} \label{lit_review}

\subsection{Motivations in OSS}
Motivations in the software industry may not apply to OSS as the contributors are independent and driven by personal motivation \citep{roberts2006understanding}. Consequently, numerous studies have explored the distinct motivational factors that drive individuals to contribute to OSS projects. \citet{ye2003toward} emphasized learning as a key intrinsic motivator. They recommended using a modular system design with independent tasks of increasing complexity, allowing \texttt{newcomers to OSS} to start small and gradually take on bigger roles. \citet{krishnamurthy2006intrinsic} identified key factors such as financial incentives, task nature, and group dynamics, noting that many contributors contribute to solving personal problems. However, \citet{hartman2011intrinsic} noted that compensation can reduce autonomy and intrinsic motivation due to social pressures. 

\citet{wu2007empirical} found that OSS contributors' long-term commitment is driven by both altruism and financial incentives, though not all contributors benefit equally. \citet{bitzer2007intrinsic} studied intrinsic motivations and highlighted the role of young, talented programmers who enjoy programming and embrace OSS’s gift culture. They emphasized the need for more research on how different motives interact and overlap among OSS contributors. \citet{carvalho2017people} categorized OSS contributors into four groups: Professionals, Specialists, Focused Labourers, and Code Enthusiasts, revealing key differences in their intrinsic motivations. However, the study had only 116 responses with limited geographic diversity, where they only had 3 responses each for the African and Oceania continents. Hence, it raises concerns about the generalizability of the findings to the broader OSS community. Additionally, it did not explore how these motivational differences translate into practical implications for project recommendation systems or contributor engagement strategies. In contrast, our work provides a more detailed analysis by explicitly examining the relationship between a comprehensive set of demographic variables and motivations, and further connecting these demographic effects to contributors’ project characteristic preferences.

Leadership as a motivator has also been explored in the OSS literature \citep{li2012leadership, wynn2004leadership}, and results revealed that transformational leadership enhances intrinsic motivation, while participative leadership fosters engagement, and achievement-oriented leadership promotes appreciation. \citet{von2012carrots} identified five dimensions of impact on motivations: licensing constraints, community sponsorship, governance, rewards, and social/technical exposure. The impact of many parameters, including the type of license on the output per contributor
in OSS projects, was examined by \citet{fershtman2007open}. \citet{gerosa2021shifting} further studied the motivations reported by \citep{lakhani2005hackers,alexander2002working,ghosh2002free}: Ideology, Altruism, Fun, Kinship, Reputation, Reciprocity, Learning, Own-use, Career, and Pay. Further, they focused on how motivations have shifted over time as contributors gain experience and highlighted demographic variations. However, they examined demographics broadly without deep exploration.

\textcolor{black}{While previous studies provide valuable insights into OSS motivations, they often focus on broad trends and do not examine how motivations relate to contributors’ project-selection preferences. Furthermore, limited attention has been paid to understanding how demographic factors shape these motivations. A deeper understanding of these correlations is necessary to explain how contributors choose OSS projects and how OSS communities can better support diverse contributor groups.}


\subsection{\textcolor{black}{OSS Project Selection and Recommendation Systems}}

\textcolor{black}{Selecting an OSS project is an important decision for both \texttt{newcomers} and \texttt{OSS practitioners}. Choosing a project that aligns with contributors’ goals, expectations, and interests can facilitate onboarding and long-term participation, whereas poor project fit may contribute to disengagement and contributor turnover \citep{phatangare2024codecompass,santos2024software}.}

\textcolor{black}{Prior research has identified several project characteristics that influence contributors' decisions when selecting OSS projects. For example, project popularity, domain, and skills required that can be extracted from project attributes, such as title, description, and ReadMe file, have been identified as important factors considered by contributors when evaluating OSS projects \citep{xu2023personalized,liao2023graph,sayce2022recommendation,ford2022reboc}. These studies primarily focus on project-level attributes that may influence contributors' project-selection decisions.}

\textcolor{black}{To support contributors in identifying suitable OSS projects, several recommendation systems have been proposed. These systems commonly utilize factors such as historical behaviour sequences of contributors \citep{shen2024multi,liao2023graph,sayce2022recommendation,xu2023personalized} in the form of commenting, forking, starring, and watching a repository, as well as skills \citep{shen2024multi,phatangare2024codecompass,xu2023personalized}.} Moreover, \citet{abhinav2023crowdassist} provided a limited exploration of motivation in OSS task recommendation systems, focusing primarily on task-based learning opportunities. Their approach evaluated whether a task aligns with a worker's motivation by assessing its potential for skill development, particularly in high-demand, high-paying skills. They used task attributes to determine job learning opportunities and introduced an algorithm to identify these skills. However, this narrow perspective did not comprehensively address the diverse and complex motivations that influence OSS contributors' project preferences.

Moreover, the recommendation systems introduced by \citet{sayce2022recommendation} and \citet{abhinav2020tasrec} considered factors such as programming language, task difficulty, and project popularity for their recommendations. However, they mentioned integrating motivation as a future enhancement to their systems. \textcolor{black}{A more in-depth understanding of contributors' motivations, demographic backgrounds, and project characteristic preferences is needed to explain how OSS contributors select projects and what factors influence these decisions. Such insights may also inform the design of future recommendation systems. }

\textcolor{black}{To address this gap, our survey investigates contributors' motivations for joining OSS projects, demographic backgrounds, and project characteristic preferences. By examining these correlations, we aim to develop a deeper understanding of OSS project-selection behaviour. Additionally, we explore contributors' perspectives on how existing recommendation systems could better support project selection within OSS communities.}

\section{Methodology} \label{method}

In this section, we introduce our Research Questions (RQs) and outline our methodology. To address our RQs, we deployed an online survey targeting individuals interested in OSS. The materials necessary to replicate this work, including the full survey questionnaire, are available in \citep{anonymous2025motivation}.

\subsection{Research Questions} \label{research_questions}

\begin{itemize}
    
   \item  \emph{\textbf{RQ1}: \textcolor{black}{How are the demographics of OSS contributors correlated with their motivations for contributing to OSS?}}

This RQ aims to provide a comprehensive analysis of the \textcolor{black}{correlations between demographic factors} (e.g., age, gender, professional experience) and the motivations of OSS contributors. \textcolor{black}{The findings of this RQ provide insight into how contributors with different demographic backgrounds report different motivational preferences, which may help inform more inclusive recommendation approaches.}

   \item   \emph{\textbf{RQ2}: \textcolor{black}{How are motivations correlated with contributors’ preferences for project characteristics?}}

\textcolor{black}{This RQ examines how contributors with different motivations report different preferences for project characteristics. The results reveal project characteristic preferences associated with each motivation, which may help inform recommendation outputs based on contributors’ stated motivations.}

  \item   \emph{\textbf{RQ3}: \textcolor{black}{How are the demographics of OSS contributors correlated with their preferences for project characteristics?}}

\textcolor{black}{This RQ aims to provide a comprehensive analysis of the correlations between demographic factors and contributors’ preferences for project characteristics. The findings of this RQ provide insight into how contributors from different demographic groups report varying project characteristic preferences.}

    \item   \emph{\textbf{RQ4}: What are the OSS contributors' views on improving existing recommendation systems?}

    This RQ captures practitioners’ perspectives on improving existing recommendation systems, identifying which areas should be prioritized and how systems can better align with contributor needs.

\end{itemize}
\subsection{Survey Design} We created an online survey incorporating both qualitative and quantitative questions, adhering to the guidelines established by Kitchenham and Pfleeger \citep{kitchenham2008personal}.  The survey was conducted in a completely anonymous manner, ensuring that participants' identities were never traced. This study received ethics approval from the relevant institutional ethics committee. Participants were provided with a clear explanation of the purpose and relevant details in the survey preamble. The advertised time to complete the survey was 10 - 15 minutes. The average time taken by participants who completed the entire survey was approximately 10 minutes. 

After reviewing the literature, we structured the survey as; screening questions (Section \ref{screening}), demographic information and motivations for joining OSS projects, preferences of project characteristics (e.g., \citep{sbai2018exploring,hata2015characteristics,fagerholm2014role}), and an open-ended question to explore their views on how project recommendation systems can be improved to better align with motivations.  Participants were allowed to select ‘Prefer not to say' for questions about gender identity and region. All questions except the open-ended one were mandatory to answer. The survey comprised a total of 30 questions, including 4 multiple-choice questions in the screening section, 10 multiple-choice questions in the demographic section, 7 Likert-scale questions in the motivations section, 4 multiple-choice and 4 Likert-scale questions in the project characteristics section, and one free-text answer space in the open-ended section.

Firstly, we recruited six practitioners for the pilot survey through personal contacts. They pointed out some minor unclear wordings in the questions, which we revised for clarity in the final survey. 

\subsubsection{Survey Questionnaire}

As mentioned in Section \ref{lit_review}, prior research has explored the motivations for joining OSS projects. \textcolor{black}{As this study focuses on contributors’ project-selection preferences, we selected seven highly cited OSS motivations from prior literature to examine how contributors with different motivations reported different preferences for OSS project characteristics during project selection.} These motivations are mentioned along with abbreviated terms in brackets that will be used throughout the remainder of the paper for clarity and consistency. \textit{Learning or gaining new skills (Learning)} where contributors seek to expand their skills \citep{gerosa2021shifting, huang2021leaving, ke2009motivations, wu2007empirical, roberts2006understanding, lakhani2005hackers, hertel2003motivation}, \textit{Helping the programmer community (Helping)} reflects the desire to help the community \citep{gerosa2021shifting, huang2021leaving, carvalho2017people, ke2009motivations, lakhani2005hackers, wu2007empirical, bitzer2007intrinsic}, and \textit{Enjoyment} as contributors find the activity inherently interesting \citep{gerosa2021shifting,huang2021leaving,ke2009motivations, bitzer2007intrinsic,roberts2006understanding,lakhani2005hackers}. \textit{Career advancement (Career)} as an opportunity to improve their employability \citep{gerosa2021shifting,huang2021leaving, ke2009motivations, wu2007empirical,roberts2006understanding,lakhani2005hackers,hertel2003motivation}, \textit{Networking and socializing (Networking)} for building connections \citep{carvalho2017people,von2012carrots,ke2009motivations,hertel2003motivation}, \textit{Gaining a reputation (Reputation)} which promotes credibility \citep{gerosa2021shifting, carvalho2017people,ke2009motivations,roberts2006understanding,lakhani2005hackers,hertel2003motivation}, and \textit{Financial incentives (Incentives)} where contributors are motivated by compensation \citep{gerosa2021shifting,huang2021leaving,von2012carrots,alexy2011fistful, roberts2006understanding,lakhani2005hackers,lerner2002some}.

 We selected a set of popular project characteristics that prior literature has identified as important factors considered by potential contributors when choosing OSS projects, and as indicators of sustainable OSS projects. \textit{Project Age} and \textit{Stage of Development} signal the maturity level of the project \citep{qiu2019signals, sbai2018exploring, fagerholm2014role}. The \textit{number of Contributors} and \textit{Followers} signal the size of the community and the support they have \citep{qiu2019signals,sbai2018exploring}. Providing \textit{Clear Guidelines} is a decisive factor, as their absence often creates an immediate negative impression \citep{fronchetti2023contributing, qiu2019signals, sbai2018exploring}. \textit{Having a Web Page} which provides more details and demonstrates the components, is considered an important factor when choosing a project \citep{qiu2019signals}. Providing \textit{Multilingual Documentation} promotes inclusiveness within the community  \citep{qiu2019signals}. Providing \textit{Source Code Comments} helps \texttt{newcomers to OSS} to understand the codebase quickly \citep{,sbai2018exploring} \textcolor{black}{ \textcolor{black}{(See Table \ref{tab:survey_questions} in Appendix \ref{tab:survey_questions} for survey questionnaire).}}

\subsubsection{Screening Process} \label{screening} Screening questions were implemented at the beginning of the survey to ensure that only eligible participants with appropriate experience and knowledge proceeded further. The screening process consisted of four mandatory questions designed to assess participants' programming experience and technical expertise.  These questions were adopted from the study \citep{danilova2021you}, which has been widely utilized in software practitioners' recruitment process \citep{russo2024navigating,tahaei2023stuck,wurzel2023competencies}. Of the 16 questions introduced by  \citet{danilova2021you}, we selected Q1, which comes in two sub-questions, Q15 and Q16, to prevent overwhelming the participants while ensuring an effective screening process. The questions included evaluating familiarity with lesser-known programming languages, interpreting a provided code fragment to identify the parameter of a function, and selecting the correct return value of the pseudocode (\textcolor{black}{See Figure \ref{fig:screening} in Appendix \ref{Appendix-a} for screening questions}).
Participants were required to answer all screening questions correctly to proceed. This approach ensured the reliability of the responses, reducing the risk of responses from fraudulent online survey takers. We did not use OSS experience as a screening criterion, as that would have excluded genuine newcomers; instead, OSS experience was captured in the demographic section.

\subsection{Participants} 
The survey participants were recruited through multiple platforms to ensure a diverse and representative dataset. \textcolor{black}{Our target population comprised software practitioners with a software engineering background, including both newcomers with no prior OSS contribution experience and experienced OSS contributors. We employed a non-probability convenience sampling strategy, which is a widely used approach in software engineering survey research. \citep{kitchenham2008personal,baltes2022sampling}} 

The invitations were sent out via Prolific \citep{prolific2025} (an online participant recruitment platform) and social media platforms like LinkedIn \citep{linkedin2025}. Prolific has been widely utilized in the software engineering research community \citep{russo2024navigating,tahaei2023stuck,reid2022software}. The Prolific platform was utilized in two recruitment rounds. We conducted recruitment in two phases: the first round without Prolific’s built-in filters, and the second round with the ‘software industry’ work-industry filter enabled to increase efficiency in targeting suitable participants. Responses were classified as ineligible during screening if participants failed one or more of the mandatory screening questions (Section \ref{screening}).
In the first round, 150 invitations were sent, and only 33 participants successfully passed the screening questions and completed the survey. The second round involved 260 invitations, and 143 completed responses were obtained. The participants from Prolific were compensated with a monetary payment accordingly.

Additionally, recruitment was conducted through social media platforms like LinkedIn and personal contacts. Ultimately, a total of 35 responses were collected via social media.

A total of 632 responses were recorded in Qualtrics. During the screening process, 417 participants (from all recruitment platforms) were deemed ineligible. Further, 4 incomplete responses were excluded. This resulted in removing 421 responses (66.6\%), leaving a final dataset of 211 completed responses. We further removed 3 (0.5\%) responses that selected ``prefer not to say" for the gender identity and region questions during the analysis. Hence, the final analysis included 208 responses (32.9\%). 

The participants in our survey had diverse demographic backgrounds. Age-wise, participants were in 18-24 \textit{(n=48)}, 25-34 \textit{(n=105)}, and 35 and over \textit{(n=55)} categories. Region-wise, Africa \textit{(n=22)}, the Americas\textit{ (n=50)}, Asia \textit{(n=43)}, Europe\textit{ (n=81)}, and Oceania \textit{(n=12)}. A complete analysis of the demographic representation of the participants is available in Table \ref{tab:demographics}.

\begin{table}[t]
\small
\caption{Demographic Factor Categories}
\label{tab:demographics}
\begin{tabularx}{\linewidth}{@{}lX@{}}
\toprule
\textbf{Factor} & \textbf{Merged Categories} \\
\midrule
Age & 18--24 (48), 25--34 (105), 35+ (55) \\
\midrule
Gender & Female (40), Male (163), Other (5) \\
\midrule
Region & Africa (22), Americas (50), Asia (43), Europe (81), Oceania (12) \\
\midrule
Education & Basic (36), Undergrad (124), Postgrad (48) \\
\midrule
Type & Non-professional (28), Professional (180) \\
\midrule
SE Experience & Novice (14), Junior (94), Senior (100) \\
\midrule
SE Role & Development (125), Data \& AI Engineer (27), Infrastructure \& Operations (22), Other (20) \\
\bottomrule
\end{tabularx}
\end{table}

\subsection{Analysis} \label{method_analysis}

For data analysis, we employed a combination of statistical analysis and thematic analysis. We used standard statistical analysis techniques to analyze the collected quantitative data. This adheres to the best practices outlined by Kitchenham and Pfleeger for analysing survey data \citep{kitchenham2008personal}. 

\textcolor{black}{Similar to prior software engineering survey studies \citep{sultana2026role,gerosa2021shifting}, some demographic categories contained comparatively fewer respondents despite our recruitment efforts across multiple platforms. To improve interpretability and reduce sparsity during statistical analysis, several categories were merged into higher-level groups. Moreover, response quality was addressed through mandatory screening questions, exclusion of incomplete responses, and removal of responses with insufficient information for qualitative analysis.}

We employed the Kruskal-Wallis \textcolor{black}{omnibus} test to examine the significance of differences across motivation preferences and demographic categories in RQ1 and project characteristic categories in RQ2. The Kruskal-Wallis test \citep{mckight2010kruskal} is a non-parametric method that does not require the assumption of normal distribution and is an extension of the Mann-Whitney U test \citep{mcknight2010mann} for comparing more than two independent groups. It evaluates whether three or more sample groups originate from the same distribution on a variable of interest. It has been widely used in software engineering research  \citep{meem2024exploring, aldndni2024understanding, ouyang2025empirical}. \textcolor{black}{Hence, we selected it as the survey responses were primarily collected using ordinal Likert-scale data and the demographic groups contained unequal sample sizes.} If the obtained p-value $\leq 0.05$, we reject the null hypothesis, concluding that there is a statistically significant difference in the mean ranks across the groups \citep{aldndni2024understanding}.

To analyze qualitative data gathered through the open-ended question, we performed \textcolor{black}{inductive} thematic analysis. It involves familiarizing with data, generating initial codes, searching for themes, reviewing themes and defining and naming themes \citep{braun2006using}. First, the first author read all the responses to become familiar with the data. Then, the responses were segmented into meaningful units, which resulted in 186 units in total. These units were subsequently coded and categorized into themes. The second and third authors independently reviewed the codes and themes, each analyzing half of the dataset. They documented any disagreements or concerns regarding the themes. Any discrepancies were resolved through multiple discussions, leading to a final set of themes agreed upon by all authors. During this process, 30 responses were excluded due to insufficient meaningful content or \textcolor{black}{because the responses were unrelated to the research question}. Following mediation, Cohen’s Kappa was calculated separately for the two halves to evaluate coding reliability, yielding scores of 0.96 and 0.83 \citep{tahaei2023stuck}. These results demonstrate a strong inter-annotator agreement, reinforcing the reliability of the coding process and the integrity of the qualitative data.

\section{Results} \label{results}

In this section, we present the analysis of our results and answers to each of the research questions. We present the results for each RQ separately for \texttt{newcomers to OSS} and experienced practitioners, as these groups may differ in their perspectives, needs, and levels of familiarity with OSS environments. This separation enables us to uncover group-specific trends, providing more nuanced insights to support personalized recommendation strategies.

\subsection{\textbf{Demographics influence on motivations (RQ1)}} \label{Rq2_1}

The demographic information of the participants was collected using multiple-choice questions. Some demographic response options were combined for data analysis due to insufficient data in certain categories, ensuring clearer and more interpretable results \citep{kitchenham2008personal,gerosa2021shifting}. Respondents were required to provide the SE role and OSS role only if they had prior experience in those fields.

In terms of the highest rated motivations among \texttt{newcomers to OSS}, \textit{Learning} was the most frequently reported motivation across nearly all demographic groups. \textit{Career} advancement also appeared prominently, especially among professionals and seniors in the SE industry. \textit{Helping} the community was particularly strong among respondents with basic education and those identifying in the “Other” gender category. Regionally, newcomers from Africa, Asia and Oceania placed more emphasis on \textit{Learning}, while those from the Americas and Europe leaned towards \textit{Career}.

In terms of the highest rated motivations among experienced \texttt{OSS practitioners}, \textit{Learning} remained important but was often accompanied by other motivations such as Career, Enjoyment, or Helping, depending on demographic subgroup. \textit{Career} was most prominent among the 18–24 age group. \textit{Helping} was highly valued in the Americas and Africa. Moreover, it was valued by the respondents with basic education, those identifying in the “Other” gender category, Data \& AI engineers, those who are in Infrastructure \& Operations roles and non-professionals.

\begin{figure*}
    \centering
    \includegraphics[width=\textwidth]{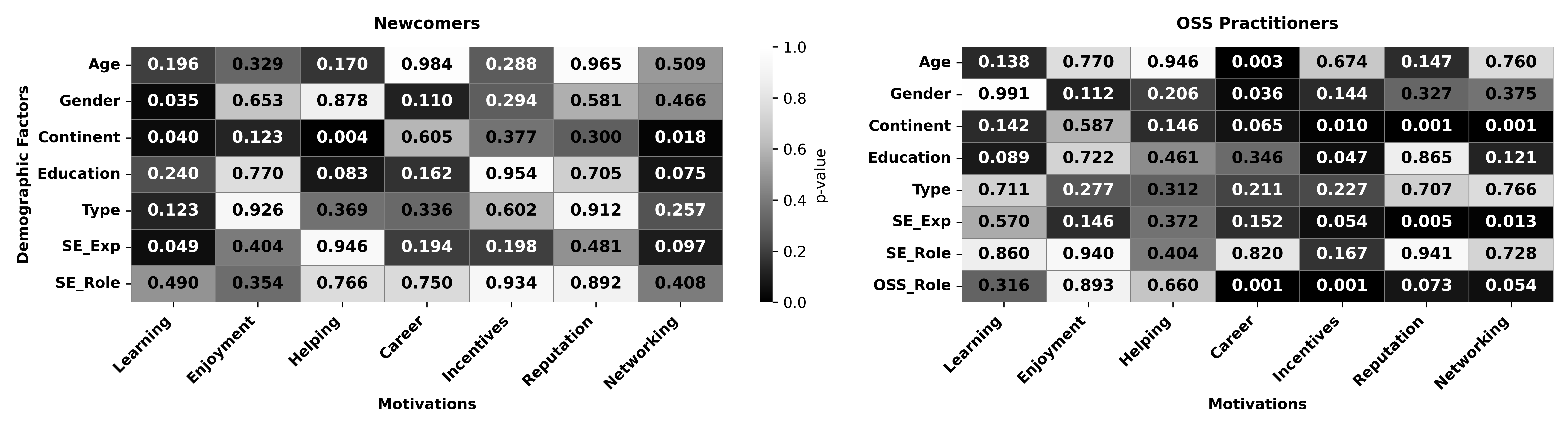}
    \caption{Motivations Vs. Demographic Factors Kruskal-Wallis \textit{p-values} Heat Maps}
    \label{demo_heatmap}
\end{figure*}

To explore statistically significant differences between the motivations and demographics categories, we performed the \textit{Kruskal-Wallis test} \citep{mckight2010kruskal} followed by Post-Hoc Bonferroni correction. All findings presented in this paper were considered statistically significant, with Bonferroni adjusted p-values less than 0.05. \textcolor{black}{Kruskal-Wallis test's p-value heat maps for Newcomers' and \texttt{OSS practitioners}' motivations and demographic factors are presented in Figure \ref{demo_heatmap}.}

 Based on the results, statistically significant differences were observed between certain demographic categories and motivation rankings. Among \texttt{newcomers to OSS}, significant differences were found for \textit{gender, region,} and \textit{SE experience}. In contrast, for experienced \texttt{OSS practitioners}, significant differences were observed for \textit{age, gender, region, education, SE experience}, and \textit{OSS role}. Then, we conducted a comprehensive analysis to examine the identified differences (data available in replication package \citep{anonymous2025motivation}). 

\subsubsection{\texttt{Newcomers}' demographic \textcolor{black}{correlations} (\textbf{n=85})}

For \textit{Learning}, Bonferroni corrected p-value of 0.038 indicated that the mean rank of females and males is significantly different. 37\% more female respondents reported high (rated as ``Very”/``Extremely”) motivation for it compared to males. Further, the p-value of 0.033 indicated a significant difference between Africa and America, with 65\% more participants from Africa rating it as having high motivation compared to those in America. Novices to the SE industry demonstrated a 17\% higher interest compared to seniors (p-value 0.045).

For \textit{Helping}, the p-value of 0.033 indicated a significant difference in mean rankings for Europe and America, where 34\% more participants from Europe rated it higher compared to those in America \textcolor{black}{(See Table \ref{tab:likert_learning_newcomers} for a detailed breakdown of the Likert-scale preference distributions).}

\textcolor{black}{Table \ref{tab:newc-stat} shows the effect size analysis for above mentioned statistically significant correlations. \textit{Learning} demonstrated a small effect for Gender ($\epsilon^2 = 0.058$) and SE experience ($\epsilon^2 = 0.049$), while the relationship between \textit{Learning} and Region showed a medium effect ($\epsilon^2 = 0.076$). In contrast, \textit{Helping} and Region exhibited a large effect size ($\epsilon^2 = 0.141$), suggesting that regional differences had a stronger practical influence on Helping motivation among newcomers to OSS.}

\begin{table*}[!htbp]
\centering
\small
\renewcommand{\arraystretch}{1.2}
\setlength{\tabcolsep}{7pt}

\caption{Likert distributions for preferences for motivations (Newcomers). Percentages in bold show “Very/Extremely” (left) and “Not at all/Slightly” (right).}
\label{tab:likert_learning_newcomers}

\newcolumntype{L}[1]{>{\raggedright\arraybackslash}p{#1}}
\newcolumntype{C}[1]{>{\centering\arraybackslash}p{#1}}

\begin{tabular}{L{0.15\textwidth} L{0.15\textwidth} | C{0.60\textwidth}}
\toprule
\textbf{Motivation} & \textbf{Factor} & \textbf{Distribution}\\
\midrule

\multirow{14}{*}{\textbf{Learning}}
& \textit{A. Gender} & \\[-1pt]
& Female & \progressbar{33.333}{46.667}{6.667}{13.333}{0}{\textbf{80\%}}{\textbf{13\%}}\\
& Male   & \progressbar{11.765}{30.882}{33.824}{20.588}{2.941}{\textbf{42\%}}{\textbf{24\%}}\\
& Other  & \progressbar{0}{0}{100}{0}{0}{\textbf{0\%}}{\textbf{0\%}}\\

\cmidrule(lr){2-3}

& \textit{B. SE Experience} & \\[-1pt]
& Newcomer & \progressbar{44.44}{33.33}{11.11}{11.11}{0}{\textbf{78\%}}{\textbf{11\%}}\\
& Junior   & \progressbar{14.29}{38.10}{26.19}{19.05}{2.38}{\textbf{52\%}}{\textbf{21\%}}\\
& Senior   & \progressbar{8.82}{26.47}{41.18}{20.59}{2.94}{\textbf{35\%}}{\textbf{24\%}}\\

\cmidrule(lr){2-3}

& \textit{C. Region} & \\[-1pt]
& Africa  & \progressbar{50}{50}{0}{0}{0}{\textbf{ 100\%}}{\textbf{0\%}}\\
& America & \progressbar{15}{20}{30}{30}{5}{\textbf{35\%}}{\textbf{35\%}}\\
& Asia    & \progressbar{16.667}{38.889}{27.778}{16.667}{0}{\textbf{56\%}}{\textbf{17\%}}\\
& Europe  & \progressbar{11.111}{36.111}{33.333}{16.667}{2.778}{\textbf{47\%}}{\textbf{19\%}}\\
& Oceania & \progressbar{0}{20}{60}{20}{0}{\textbf{20\%}}{\textbf{20\%}}\\

\midrule
\multirow{7}{*}{\textbf{Helping}}
& \textit{A. Region} & \\[-1pt]
& Africa  & \progressbar{0}{33.333}{33.333}{33.333}{0}{\textbf{33\%}}{\textbf{33\%}} \\
& America & \progressbar{0}{10}{25}{30}{35}{\textbf{10\%}}{\textbf{65\%}} \\
& Asia    & \progressbar{0}{44.444}{22.222}{27.778}{5.556}{\textbf{44\%}}{\textbf{33\%}} \\
& Europe  & \progressbar{5.556}{38.889}{30.556}{19.444}{5.556}{\textbf{44\%}}{\textbf{25\%}} \\
& Oceania & \progressbar{0}{0}{40}{40}{20}{\textbf{0\%}}{\textbf{60\%}} \\
\bottomrule

\end{tabular}

\vspace{0.6em}
\likertlegend
\end{table*}

\begin{table}[htbp]
\caption{Newcomers' motivations and demographics statistics}
\label{tab:newc-stat}
\centering
\begin{tabular}{@{}llllllll@{}}
\toprule
Motivation & Demographic Category & H & df & p & k & $\epsilon^2$ & Effect \\ 
\midrule
Learning & Gender & 6.715 & 2 & 0.035 & 3 & 0.058 & Small \\
 & Region & 10.04 & 4 & 0.04 & 5 & 0.076 & Medium \\
 & SE experience & 6.022 & 2 & 0.049 & 3 & 0.049 & Small \\
Helping & Region & 15.318 & 4 & 0.004 & 5 & 0.141 & Large \\
\bottomrule
\end{tabular}
\end{table}

\subsubsection{\texttt{OSS practitioners}' demographic \textcolor{black}{correlations}  (\textbf{n=123})}

For \textit{Career}, 36\% more respondents aged 18–24 rated it highly compared to those aged 35 and above (p-value 0.005). Additionally, 17\% more individuals in the 25-34 age group rated it higher than the group aged 35 and above (p-value 0.029). p-values of 0.045 and 0.03 showed significant mean ranking differences between the Female and Other categories and between the Male and Other categories, respectively. All respondents from the Other category rated it as a low (``Slighly”/``Not at all”). Whereas 56\% of Females and 49\% of Males rated it as high. 15\% more core contributors to OSS rated it as having high motivation compared to casual contributors (p-value 0.025). Moreover, 17\% more core contributors to OSS rated it as having high motivation compared to peripheral contributors (p-value 0.002).

For \textit{Incentives}, 33\% more respondents from Africa rated it highly compared to those from America (p-value 0.011). Moreover, 34\% more respondents from Africa rated it highly compared to those from Europe (p-value 0.01). 7\% more respondents with Postgraduate qualifications rated it higher than those with basic education qualifications (p-value 0.042). However, those who rated it as low motivation compared to high are higher for both education qualification categories.
Seniors in the SE industry demonstrated a 23\% higher interest compared to novices (p-value 0.047). Moreover, all novices rate it as low motivation. Core contributors to OSS are 13\% more likely to rate their motivation as high compared to both casual and peripheral contributors, with a statistically significant p-value of 0.001.

For \textit{Reputation}, 56\% more respondents from Africa rated it highly compared to those from America (p-value 0.001). Moreover, 38\% more respondents from Africa rated it highly compared to those from Europe (p-value 0.01). 30\% more juniors in the SE industry rated it as having higher motivation compared to seniors (p-value = 0.006).

For \textit{Networking}, 29\% more respondents from Africa rated it highly compared to those from America (p-value 0.003). Additionally, 47\% more respondents from Africa rated it highly compared to those from Europe (p-value 0.002). 19\% more juniors in the SE industry rated it as having higher motivation compared to seniors (p-value = 0.026) \textcolor{black}{(See Tables \ref{tab:career_op} and \ref{tab:likert_op_continued} for a detailed breakdown of the Likert-scale preference distributions),}

\textcolor{black}{
The effect size analysis (See Table \ref{pract-demo-stat}) further revealed varying levels of practical significance across the identified correlations. While several correlations demonstrated small-to-medium effects, Reputation ($\epsilon^2 = 0.158$) and Networking ($\epsilon^2 = 0.148$) with Region exhibited large effect sizes, indicating that geographical differences had a substantial influence on these motivations among OSS practitioners.
}

\begin{table*}[!htbp]
\centering
\small
\renewcommand{\arraystretch}{1.2}
\setlength{\tabcolsep}{7pt}

\caption{Likert distributions for preference for Motivations (OSS Practitioners). Percentages in bold show “Very/Extremely” (left) and “Not at all/Slightly” (right).}
\label{tab:career_op}

\newcolumntype{L}[1]{>{\raggedright\arraybackslash}p{#1}}
\newcolumntype{C}[1]{>{\centering\arraybackslash}p{#1}}

\begin{tabular}{L{0.15\textwidth} L{0.15\textwidth} | C{0.60\textwidth}}
\toprule
\textbf{Motivation} & \textbf{Factor} & \textbf{Distribution}\\
\midrule

\multirow{11}{*}{\textbf{Career}}
& \textit{A. Age} & \\[-1pt]
& 18--24 & \progressbar{37.037}{33.333}{14.815}{7.407}{7.407}{\textbf{ 70\%}}{\textbf{15\%}}\\
& 25--34 & \progressbar{32.727}{18.182}{29.091}{14.545}{5.455}{\textbf{ 51\%}}{\textbf{20\%}}\\
& 35+    & \progressbar{7.317}{26.829}{29.268}{21.951}{14.634}{\textbf{ 34\%}}{\textbf{37\%}}\\

\cmidrule(lr){2-3}

& \textit{B. Gender} & \\[-1pt]
& Female & \progressbar{24}{32}{20}{8}{16}{\textbf{ 56\%}}{\textbf{24\%}}\\
& Male   & \progressbar{26.316}{23.158}{28.421}{16.842}{5.263}{\textbf{ 49\%}}{\textbf{22\%}}\\
& Other  & \progressbar{0}{0}{0}{33.333}{66.667}{\textbf{0\%}}{\textbf{100\%}}\\

\cmidrule(lr){2-3}

& \textit{C. OSS Role} & \\[-1pt]
& Core       & \progressbar{27.193}{35.088}{27.193}{7.018}{3.509}{\textbf{ 62\%}}{\textbf{11\%}}\\
& Peripheral & \progressbar{20.290}{14.493}{30.435}{21.739}{13.043}{\textbf{ 35\%}}{\textbf{35\%}}\\
& Casual     & \progressbar{25.806}{21.505}{23.656}{18.280}{10.753}{\textbf{ 47\%}}{\textbf{29\%}}\\

\midrule

\multirow{18}{*}{\textbf{Incentives}}
& \textit{A. Education} & \\[-1pt]
& Basic         & \progressbar{4.167}{4.167}{8.333}{20.833}{62.5}{\textbf{ 8\%}}{\textbf{83\%}}\\
& Undergraduate & \progressbar{14.286}{12.857}{15.714}{20}{37.143}{\textbf{ 27\%}}{\textbf{57\%}}\\
& Postgraduate  & \progressbar{13.793}{3.448}{27.586}{13.793}{41.379}{\textbf{ 17\%}}{\textbf{55\%}}\\

\cmidrule(lr){2-3}

& \textit{B. SE Experience} & \\[-1pt]
& Newcomer & \progressbar{0}{0}{0}{0}{100}{\textbf{0\%}}{\textbf{100\%}}\\
& Junior   & \progressbar{11.538}{9.615}{17.308}{17.308}{44.231}{\textbf{ 21\%}}{\textbf{62\%}}\\
& Senior   & \progressbar{13.636}{9.091}{18.182}{21.212}{37.879}{\textbf{ 23\%}}{\textbf{59\%}}\\

\cmidrule(lr){2-3}

& \textit{C. Region} & \\[-1pt]
& Africa  & \progressbar{37.5}{12.5}{18.75}{18.75}{12.5}{\textbf{ 50\%}}{\textbf{31\%}}\\
& America & \progressbar{10}{6.667}{13.333}{13.333}{56.667}{\textbf{ 17\%}}{\textbf{70\%}}\\
& Asia    & \progressbar{4}{16}{32}{4}{44}{\textbf{ 20\%}}{\textbf{48\%}}\\
& Europe  & \progressbar{11.111}{4.444}{8.889}{26.667}{48.889}{\textbf{ 16\%}}{\textbf{76\%}}\\
& Oceania & \progressbar{0}{14.286}{28.571}{42.857}{14.286}{\textbf{ 14\%}}{\textbf{57\%}}\\

\cmidrule(lr){2-3}

& \textit{D. OSS Role} & \\[-1pt]
& Core       & \progressbar{18.421}{11.404}{31.579}{10.526}{28.070}{\textbf{ 30\%}}{\textbf{39\%}}\\
& Peripheral & \progressbar{8.696}{8.696}{7.246}{24.638}{50.725}{\textbf{ 17\%}}{\textbf{75\%}}\\
& Casual     & \progressbar{9.677}{7.527}{11.828}{21.505}{49.462}{\textbf{ 17\%}}{\textbf{71\%}}\\

\bottomrule
\end{tabular}

\vspace{0.6em}
\likertlegend
\end{table*}

\begin{table*}[!htbp]
\centering
\small
\renewcommand{\arraystretch}{1.2}
\setlength{\tabcolsep}{7pt}

\caption{Cont. Likert distributions for preference for Motivations (OSS Practitioners). Percentages in bold show “Very/Extremely” (left) and “Not at all/Slightly” (right)}
\label{tab:likert_op_continued}

\newcolumntype{L}[1]{>{\raggedright\arraybackslash}p{#1}}
\newcolumntype{C}[1]{>{\centering\arraybackslash}p{#1}}

\begin{tabular}{L{0.15\textwidth} L{0.15\textwidth} | C{0.60\textwidth}}
\toprule
\textbf{Motivation} & \textbf{Factor} & \textbf{Distribution}\\
\midrule

\multirow{11}{*}{\textbf{Reputation}}
& \textit{A. Region} & \\[-1pt]
& Africa  & \progressbar{37.5}{31.25}{25}{0}{6.25}{\textbf{69\%}}{\textbf{6\%}}\\
& America & \progressbar{10}{3.333}{23.333}{43.333}{20}{\textbf{13\%}}{\textbf{63\%}}\\
& Asia    & \progressbar{16}{32}{24}{20}{8}{\textbf{48\%}}{\textbf{28\%}}\\
& Europe  & \progressbar{8.889}{22.222}{17.778}{31.111}{20}{\textbf{31\%}}{\textbf{51\%}}\\
& Oceania & \progressbar{28.571}{14.286}{42.857}{14.286}{0}{\textbf{43\%}}{\textbf{14\%}}\\

\cmidrule(lr){2-3}

& \textit{B. SE Experience} & \\[-1pt]
& Newcomer & \progressbar{0}{0}{60}{20}{20}{\textbf{0\%}}{\textbf{40\%}}\\
& Junior   & \progressbar{28.846}{25}{15.385}{19.231}{11.538}{\textbf{54\%}}{\textbf{31\%}}\\
& Senior   & \progressbar{6.061}{18.182}{25.758}{33.333}{16.667}{\textbf{24\%}}{\textbf{50\%}}\\

\midrule

\multirow{11}{*}{\textbf{Networking}}
& \textit{A. Region} & \\[-1pt]
& Africa  & \progressbar{43.75}{25}{18.75}{12.5}{0}{\textbf{69\%}}{\textbf{13\%}}\\
& America & \progressbar{6.667}{23.333}{20}{20}{30}{\textbf{30\%}}{\textbf{50\%}}\\
& Asia    & \progressbar{12}{32}{32}{24}{0}{\textbf{44\%}}{\textbf{24\%}}\\
& Europe  & \progressbar{2.222}{20}{35.556}{20}{22.222}{\textbf{22\%}}{\textbf{42\%}}\\
& Oceania & \progressbar{14.286}{14.286}{42.857}{28.571}{0}{\textbf{29\%}}{\textbf{29\%}}\\

\cmidrule(lr){2-3}

& \textit{B. SE Experience} & \\[-1pt]
& Newcomer & \progressbar{0}{20}{20}{20}{40}{\textbf{20\%}}{\textbf{60\%}}\\
& Junior   & \progressbar{23.077}{23.077}{28.846}{13.462}{11.538}{\textbf{46\%}}{\textbf{25\%}}\\
& Senior   & \progressbar{3.030}{24.242}{30.303}{25.758}{16.667}{\textbf{27\%}}{\textbf{42\%}}\\

\bottomrule
\end{tabular}

\vspace{0.6em}
\likertlegend
\end{table*}

\begin{table}[htbp]
\caption{OSS Practitioners' motivations and demographics statistics}
\label{pract-demo-stat}
\centering
\begin{tabular}{@{}llllllll@{}}
\toprule
Motivation & Demographic Category & H & df & p & k & $\epsilon^2$ & Effect \\ 
Career & Age & 11.363 & 2 & 0.003 & 3 & 0.078 & Medium \\
 & Gender & 6.651 & 2 & 0.036 & 3 & 0.039 & Small \\
 & OSS role & 13.172 & 2 & 0.001 & 3 & 0.031 & Small \\
Incentives & Education & 6.131 & 2 & 0.047 & 3 & 0.034 & Small \\
 & SE Experience & 5.85 & 2 & 0.054 & 3 & 0.032 & Small \\
 & Region & 13.379 & 4 & 0.01 & 5 & 0.079 & Medium \\
 & OSS role & 24.48 & 2 & 0.001 & 3 & 0.061 & Medium \\
Reputation & Region & 22.657 & 4 & 0.001 & 5 & 0.158 & Large \\
 & SE Experience & 10.446 & 2 & 0.005 & 3 & 0.070 & Medium \\
Networking & Region & 21.421 & 4 & 0.001 & 5 & 0.148 & Large \\
 & SE Experience & 8.63 & 2 & 0.013 & 3 & 0.055 & Small \\ \bottomrule
\end{tabular}
\end{table}

\begin{tcolorbox}[arc=1mm,width=1.0\columnwidth,
                  top=1mm,left=1mm,  right=1mm, bottom=1mm,
                  boxrule=.75pt]
 \textbf{RQ 1 Summary:} \textcolor{black}{Demographic attributes showed statistically significant correlations with contributors’ motivations.} For \texttt{Newcomers to OSS}, Learning dominated across most groups, with Career and Helping varying by age, region, and gender. For \texttt{OSS practitioners}, motivations were more diverse, with Learning, Enjoyment, and Helping prominent among younger groups, those who have basic education qualifications and nonprofessionals.
\end{tcolorbox}

\subsection {\textbf{Motivations influence on project characteristics (RQ2)}} 

The preferences for 8 project characteristics were collected using multiple-choice and Likert-type questions. Some of the project characteristic response options were merged for the data analysis as in \ref{Rq2_1}.

In terms of the highest rated project characteristics for \texttt{newcomers to OSS}, \textit{new projects} (\textless 3 years old) in \textit{active development} were consistently favored across all motivations. \textit{Small contributor groups} (\textless 50 contributors) were generally preferred, except for Enjoyment, Incentives and Networking, where “it doesn’t matter” appeared as the highest response. \textit{Follower counts} were largely seen as unimportant across all motivations except for Enjoyment, where they preferred a \textit{small number of followers} (\textless 50). Having \textit{clear contribution guidelines} was rated extremely important in every case, while \textit{having their own webpage} was considered only moderately important in all cases. \textit{Multilingual documentation} was generally viewed as not at all important, with the sole exception of Networking, where it was rated extreme. Finally, \textit{having source code comments} was consistently rated either extremely or very important across all motivations.

In terms of the highest-rated project characteristics for experienced \texttt{OSS practitioners}, the same general pattern emerged: \textit{new projects} in \textit{active development} were preferred across all motivations. \textit{Small contributor groups} were valued in most cases, though Enjoyment included both “Doesn’t matter” and Small as top ratings. \textit{Follower counts} were again largely rated as ``Doesn't matter''. \textit{Clear contribution guidelines} were consistently rated as very important or extremely important, depending on motivation. \textit{Having their own webpage }was only considered moderately important. \textit{Multilingual documentation} was largely viewed as unimportant. However, it was rated as moderately important for those who have higher motivation for \textit{Incentives}. \textit{Source code comments} were rated as very important or extreme across all groups.

\begin{figure*}
    \centering
    \includegraphics[width=\textwidth]{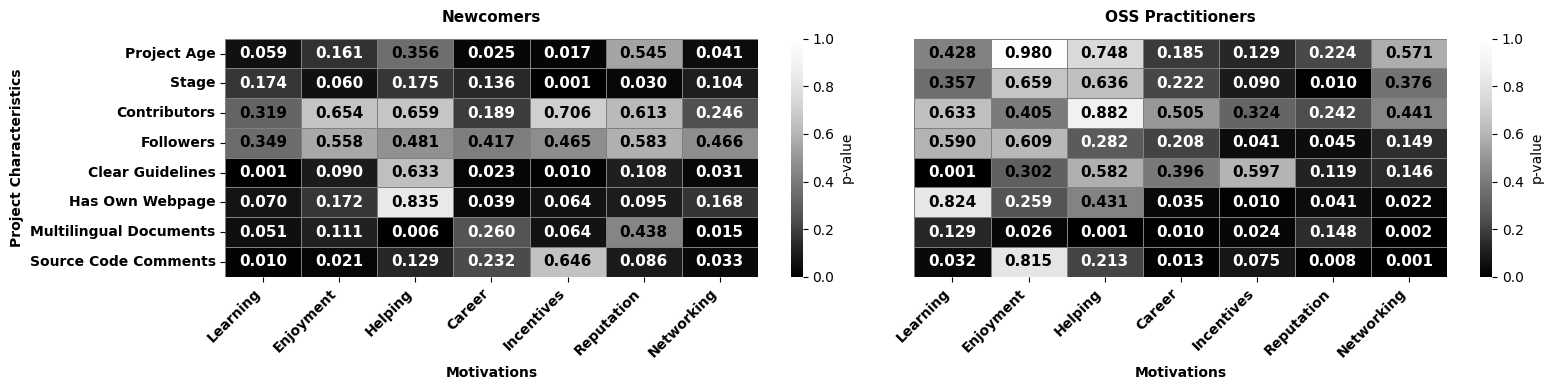}
    \caption{Motivations Vs. Project Characteristics Kruskal-Wallis \textit{p-values} Heat Maps}
    \label{proj_heatmap}
\end{figure*}

To explore the statistically significant differences between the choice of project characteristics and motivations, we performed the \textit{Kruskal-Wallis test} \citep{mckight2010kruskal} followed by Post-Hoc Bonferroni correction. All findings presented in this paper were statistically significant, with Bonferroni adjusted p-values less than 0.05. \textcolor{black}{Kruskal-Wallis test's p-value heat maps for Newcomers' and \texttt{OSS practitioners}' motivations and project characteristic preferences are presented in Figure \ref{proj_heatmap}.}

The analysis of project characteristics and motivations reveals differing patterns between \texttt{newcomers to OSS} and \texttt{OSS practitioners}. For newcomers, all project characteristics, except follower counts and contributor counts, showed statistically significant differences in their ratings. For \texttt{OSS practitioners}, all project characteristics except project age, stage and contributor counts had statistically significant differences in their ratings. Then, we comprehensively analyzed how motivation influences the selection of project characteristics (data available in replication package \citep{anonymous2025motivation}).

\subsubsection{\texttt{Newcomers}' motivational \textcolor{black}{correlations}  \textbf{(n=85)}}

For participants with high motivation for \textit{Learning}, 52\% more rated having clear contribution guidelines as extremely important compared to slightly (p-value 0.011), 42\% more compared to moderately (p-value 0.001), and 25\% more compared to very important (p-value 0.011). Moreover, 41\% more rated providing multilingual documentation as extremely important compared to not at all (p-value 0.016). For participants with high motivation for \textit{Helping}, 21\% more rated providing multilingual documentation as moderately important compared to slightly (p-value 0.005).
For participants with high motivation for \textit{Networking}, 16\% more rated providing multilingual documentation as very important compared to not at all (p-value 0.022).

For participants with high motivation for \textit{Incentives}, Bonferroni corrected p-value of 0.013 indicated that the mean rank of ``Doesn't matter'' and New project age categories is significantly different. 50\% more participants preferred New projects compared to those who rated it as ``Doesn't matter''. Moreover, 39\% more participants rated having clear contribution guidelines as extremely important compared to very (p-value 0.015) \textcolor{black}{(See Tables \ref{tab:likert_learning_projchars_nc} and \ref{tab:incentives_projchars_nc} for a detailed breakdown of the Likert-scale preference distributions).}

\textcolor{black}{
The effect size analysis in Table \ref{tab:newc-pc-stat} shows that, among newcomers to OSS, the correlation between Learning motivation and having clear contribution guidelines had the strongest practical effect ($\epsilon^2 = 0.232$), indicating a large effect. The remaining significant correlations showed medium effects.
}

\begin{table*}[!htbp]
\centering
\small
\renewcommand{\arraystretch}{1.2}
\setlength{\tabcolsep}{7pt}

\caption{Likert distributions for preferences for project characteristics (Newcomers). Percentages in bold show “Very/Extremely” (left) and “Not at all/Slightly” (right).}
\label{tab:likert_learning_projchars_nc}

\newcolumntype{L}[1]{>{\raggedright\arraybackslash}p{#1}}
\newcolumntype{C}[1]{>{\centering\arraybackslash}p{#1}}

\begin{tabular}{L{0.15\textwidth} L{0.15\textwidth} | C{0.6\textwidth}}
\toprule
\textbf{Motivation} & \textbf{Motivation Intensity} & \textbf{Distribution}\\
\midrule

\multirow{12}{*}{\textbf{Learning}}

&  & \textit{A. Having Clear Guidelines}\\[-8pt]
& Not at all  & \progressbar{0}{0}{50}{0}{50}{\textbf{0\%}}{\textbf{50\%}}\\
& Slightly    & \progressbar{6.25}{37.5}{37.5}{18.75}{0}{\textbf{44\%}}{\textbf{19\%}}\\
& Moderately  & \progressbar{23.077}{46.154}{19.231}{11.538}{0}{\textbf{69\%}}{\textbf{12\%}}\\
& Very        & \progressbar{39.286}{39.286}{17.857}{3.571}{0}{\textbf{79\%}}{\textbf{4\%}}\\
& Extremely   & \progressbar{84.615}{7.692}{0}{0}{7.692}{\textbf{92\%}}{\textbf{8\%}}\\

\cmidrule(lr){2-3}

&  & \textit{B. Multilingual Documentation}\\[-2pt]
& Not at all  & \progressbar{0}{0}{0}{0}{100}{\textbf{0\%}}{\textbf{100\%}}\\
& Slightly    & \progressbar{0}{6.25}{12.5}{18.75}{62.5}{\textbf{6\%}}{\textbf{81\%}}\\
& Moderately  & \progressbar{3.846}{7.692}{26.923}{26.923}{34.615}{\textbf{12\%}}{\textbf{62\%}}\\
& Very        & \progressbar{7.143}{14.286}{14.286}{25}{39.286}{\textbf{21\%}}{\textbf{64\%}}\\
& Extremely   & \progressbar{15.385}{23.077}{30.769}{15.385}{15.385}{\textbf{38\%}}{\textbf{31\%}}\\

\midrule

\multirow{6}{*}{\textbf{Helping}}
&  & \textit{A. Multilingual Documentation}\\[-2pt]
& Not at all  & \progressbar{9.091}{9.091}{0}{45.455}{36.364}{\textbf{18\%}}{\textbf{82\%}}\\
& Slightly    & \progressbar{9.091}{4.545}{4.545}{31.818}{50}{\textbf{14\%}}{\textbf{82\%}}\\
& Moderately  & \progressbar{4.167}{12.5}{29.167}{16.667}{37.5}{\textbf{17\%}}{\textbf{54\%}}\\
& Very        & \progressbar{3.846}{19.231}{30.769}{11.538}{34.615}{\textbf{23\%}}{\textbf{46\%}}\\
& Extremely   & \progressbar{0}{0}{50}{0}{50}{\textbf{0\%}}{\textbf{50\%}}\\

\midrule

\multirow{6}{*}{\textbf{Networking}}
&  & \textit{A. Multilingual Documentation}\\[-2pt]
& Not at all  & \progressbar{5}{5}{10}{20}{60}{\textbf{10\%}}{\textbf{80\%}}\\
& Slightly    & \progressbar{0}{6.667}{13.333}{33.333}{46.667}{\textbf{7\%}}{\textbf{80\%}}\\
& Moderately  & \progressbar{9.375}{6.25}{25}{21.875}{37.5}{\textbf{16\%}}{\textbf{59\%}}\\
& Very        & \progressbar{7.692}{30.769}{30.769}{7.692}{23.077}{\textbf{38\%}}{\textbf{31\%}}\\
& Extremely   & \progressbar{0}{7.692}{20}{40}{0}{\textbf{8\%}}{\textbf{40\%}}\\

\bottomrule
\end{tabular}

\vspace{0.6em}
\likertlegend
\end{table*}

\begin{table*}[!htbp]
\centering
\small
\renewcommand{\arraystretch}{1.2}
\setlength{\tabcolsep}{7pt}

\caption{Contd. Likert distributions for preferences for project characteristics (Newcomers). Percentages in bold show “Very/Extremely” (left) and “Not at all/Slightly” (right).}
\label{tab:incentives_projchars_nc}

\newcolumntype{L}[1]{>{\raggedright\arraybackslash}p{#1}}
\newcolumntype{C}[1]{>{\centering\arraybackslash}p{#1}}

\begin{tabular}{L{0.15\textwidth} L{0.15\textwidth} | C{0.6\textwidth}}
\toprule
\textbf{Motivation} & \textbf{Motivation Intensity} & \textbf{Distribution}\\
\midrule

\multirow{12}{*}{\textbf{Incentives}}

&  & \textit{A. Project Age}\\[-6pt]
& Not at all  & \progressbar{0}{0}{37.037}{3.704}{59.259}{\textbf{0\%}}{\textbf{63\%}}\\
& Slightly    & \progressbar{0}{0}{52.632}{5.263}{42.105}{\textbf{0\%}}{\textbf{47\%}}\\
& Moderately  & \progressbar{0}{0}{69.231}{0}{30.769}{\textbf{0\%}}{\textbf{31\%}}\\
& Very        & \progressbar{0}{0}{70}{5}{25}{\textbf{0\%}}{\textbf{30\%}}\\
& Extremely   & \progressbar{0}{0}{83.333}{0}{16.667}{\textbf{0\%}}{\textbf{17\%}}\\

\cmidrule(lr){2-3}

&  & \textit{B. Having Clear Guidelines}\\[-2pt]
& Not at all  & \progressbar{14.815}{44.444}{22.222}{14.815}{3.704}{\textbf{59\%}}{\textbf{19\%}}\\
& Slightly    & \progressbar{31.579}{47.368}{15.789}{5.263}{0}{\textbf{79\%}}{\textbf{5\%}}\\
& Moderately  & \progressbar{23.077}{23.077}{46.154}{7.692}{0}{\textbf{46\%}}{\textbf{8\%}}\\
& Very        & \progressbar{55}{30}{10}{5}{0}{\textbf{85\%}}{\textbf{5\%}}\\
& Extremely   & \progressbar{83.333}{0}{0}{0}{16.667}{\textbf{83\%}}{\textbf{17\%}}\\

\bottomrule
\end{tabular}

\vspace{0.8em}

{\footnotesize
\textbf{Project Age scale:}\quad
\legendbox{gray!85}\, Doesn't matter \quad
\legendbox{gray!55}\, Mature \quad
\legendbox{gray!25}\, New
}

\vspace{0.4em}

{\footnotesize
\textbf{Clear Guidelines scale:}\quad
\legendbox{NotAtAll}\, Not at all \quad
\legendbox{Slightly}\, Slightly \quad
\legendbox{Moderately}\, Moderately \quad
\legendbox{Very}\, Very \quad
\legendbox{Extremely}\, Extremely
}

\end{table*}

\begin{table}[htbp]
\caption{Newcomers' motivations and project characteristics statistics}
\label{tab:newc-pc-stat}
\centering
\begin{tabular}{@{}llllllll@{}}
\toprule
Motivation & Project Characteristics & H & df & p & k & $\epsilon^2$ & Effect \\ \midrule
Learning & Having clear guidelines & 22.527 & 4 & 0.001 & 5 & 0.232 & Large \\
 & Multilingual Documentation & 9.453 & 4 & 0.051 & 5 & 0.068 & Medium \\
Helping & Multilingual Documentation & 14.485 & 4 & 0.006 & 5 & 0.131 & Medium \\
Networking & Multilingual Documentation & 12.322 & 4 & 0.015 & 5 & 0.104 & Medium \\
Incentives & Project Age & 8.128 & 2 & 0.017 & 3 & 0.075 & Medium \\
 & Having clear guidelines & 13.314 & 4 & 0.01 & 5 & 0.116 & Medium \\ 
\bottomrule
\end{tabular}
\end{table}

\subsubsection{\texttt{OSS practitioners}' motivational \textcolor{black}{correlations}  \textbf{(n=123)}}

For participants with high motivation for \textit{Learning}, 31\% more rated having clear contribution guidelines as extremely important compared to slightly (p-value 0.043), 9\% more compared to moderately (p-value 0.011). For participants with high motivation for \textit{Reputation}, 30\% more rated having source code comments as extremely important compared to slightly (p-value 0.007).

For participants with high \textit{Career} motivation, 39\% more rated having their own web page as moderately important compared to not at all (p-value 0.039). Moreover, 5\% more rated providing multilingual documentation as not at all important compared to moderately (p-value 0.029). 33\% more rated having source code comments as extremely important compared to slightly (p-value 0.008).

For participants with high motivation for \textit{Incentives}, 23\% more rated having their own web page as extremely important compared to not at all (p-value 0.014). Further, 8\% more rated providing multilingual documentation as not at all important compared to moderately (p-value 0.024).
Participants with high \textit{Networking} motivation indicated that 3\% more rated having their own web page as extremely important compared to slightly (p-value 0.042). Additionally, 5\% more rated providing multilingual documentation as not at all important compared to extremely (p-value 0.003). 30\% more rated having source code comments as extremely important compared to not at all (p-value 0.005) and slightly(p-value 0.01).

For participants with high motivation for \textit{Enjoyment}, 1\% more rated providing multilingual documentation as extremely important compared to moderately (p-value 0.027). For participants with high motivation for \textit{Helping}, 32\% more rated providing multilingual documentation as not at all important compared to slightly(p-value 0.036), 1\% more as extremely important compared to slightly (p-value 0.019), 2\% more as extremely important compared to moderately (p-value 0.026) \textcolor{black}{(See Tables \ref{tab:projchars_op_learning_reputation}, \ref{tab:projchars_op_continued_incentives}, and \ref{tab:projchars_op_continued_enjoyment_helping} for a detailed breakdown of the Likert-scale preference distributions).}

\textcolor{black}{
The effect size analysis in Table \ref{tab:pract-pc-stat} indicated varying levels of practical significance across the identified correlations. Most statistically significant correlations demonstrated medium effect sizes, suggesting a moderate practical influence of motivations on project characteristic preferences among OSS practitioners. The strongest practical effects were observed for Networking with Having Source Code Comments ($\epsilon^2 = 0.129$), Helping with Having Multilingual Documentation ($\epsilon^2 = 0.123$), and Learning with Having Clear Guidelines ($\epsilon^2 = 0.120$). In contrast, Career and Having Own Web Page demonstrated a smaller practical effect ($\epsilon^2 = 0.054$), indicating a comparatively weaker effect on project selection preferences.
}
\begin{table*}[!htbp]
\centering
\small
\renewcommand{\arraystretch}{1.2}
\setlength{\tabcolsep}{7pt}

\caption{Likert distributions for preferences for project characteristics (OSS Practitioners). Percentages in bold show “Very/Extremely” (left) and “Not at all/Slightly” (right).}
\label{tab:projchars_op_learning_reputation}

\newcolumntype{L}[1]{>{\raggedright\arraybackslash}p{#1}}
\newcolumntype{C}[1]{>{\centering\arraybackslash}p{#1}}

\begin{tabular}{L{0.15\textwidth} L{0.15\textwidth} | C{0.60\textwidth}}
\toprule
\textbf{Motivation} & \textbf{Motivation Intensity} & \textbf{Distribution}\\
\midrule

\multirow{6}{*}{\textbf{Learning}}
&  & \textit{A. Having Clear Guidelines}\\[-2pt]
& Not at all  & \progressbar{0.000}{0.000}{100.000}{0.000}{0.000}{\textbf{0\%}}{\textbf{0\%}}\\
& Slightly    & \progressbar{25.000}{16.667}{25.000}{25.000}{8.333}{\textbf{42\%}}{\textbf{33\%}}\\
& Moderately  & \progressbar{7.692}{38.462}{46.154}{3.846}{3.846}{\textbf{46\%}}{\textbf{8\%}}\\
& Very        & \progressbar{20.000}{44.444}{31.111}{4.444}{0.000}{\textbf{64\%}}{\textbf{4\%}}\\
& Extremely   & \progressbar{51.282}{25.641}{20.513}{2.564}{0.000}{\textbf{77\%}}{\textbf{3\%}}\\

\midrule

\multirow{6}{*}{\textbf{Reputation}}
&  & \textit{B. Having Source Code Comments}\\[-2pt]
& Not at all  & \progressbar{11.111}{38.889}{16.667}{27.778}{5.556}{\textbf{50\%}}{\textbf{33\%}}\\
& Slightly    & \progressbar{12.121}{39.394}{30.303}{12.121}{6.061}{\textbf{52\%}}{\textbf{18\%}}\\
& Moderately  & \progressbar{32.143}{28.571}{25.000}{10.714}{3.571}{\textbf{61\%}}{\textbf{14\%}}\\
& Very        & \progressbar{16.000}{40.000}{40.000}{0.000}{4.000}{\textbf{56\%}}{\textbf{4\%}}\\
& Extremely   & \progressbar{52.632}{21.053}{21.053}{5.263}{0.000}{\textbf{74\%}}{\textbf{5\%}}\\

\midrule
\multirow{18}{*}{\textbf{Career}}

&  & \textit{A. Having Own Webpage}\\[-8pt]
& Not at all  & \progressbar{0.000}{0.000}{0.000}{45.455}{27.273}{\textbf{0\%}}{\textbf{73\%}}\\
& Slightly    & \progressbar{10.526}{10.526}{42.105}{26.316}{10.526}{\textbf{21\%}}{\textbf{37\%}}\\
& Moderately  & \progressbar{9.375}{18.750}{37.500}{25.000}{9.375}{\textbf{28\%}}{\textbf{34\%}}\\
& Very        & \progressbar{20.000}{20.000}{43.333}{13.333}{3.333}{\textbf{40\%}}{\textbf{17\%}}\\
& Extremely   & \progressbar{9.677}{25.806}{38.710}{25.806}{0.000}{\textbf{35\%}}{\textbf{26\%}}\\

\cmidrule(lr){2-3}

&  & \textit{B. Multilingual Documentation}\\[-2pt]
& Not at all  & \progressbar{0.000}{9.091}{18.182}{0.000}{72.727}{\textbf{9\%}}{\textbf{73\%}}\\
& Slightly    & \progressbar{10.526}{15.789}{0.000}{10.526}{63.158}{\textbf{26\%}}{\textbf{74\%}}\\
& Moderately  & \progressbar{3.125}{9.375}{15.625}{46.875}{25.000}{\textbf{13\%}}{\textbf{72\%}}\\
& Very        & \progressbar{13.333}{13.333}{23.333}{10.000}{40.000}{\textbf{27\%}}{\textbf{50\%}}\\
& Extremely   & \progressbar{19.355}{9.677}{29.032}{19.355}{22.581}{\textbf{29\%}}{\textbf{42\%}}\\

\cmidrule(lr){2-3}

&  & \textit{C. Having Source Code Comments}\\[-2pt]
& Not at all  & \progressbar{0}{36.364}{18.182}{27.273}{18.182}{\textbf{36\%}}{\textbf{45\%}}\\
& Slightly    & \progressbar{26.316}{26.316}{26.316}{15.789}{5.263}{\textbf{53\%}}{\textbf{21\%}}\\
& Moderately  & \progressbar{6.250}{40.625}{37.500}{15.625}{0.000}{\textbf{47\%}}{\textbf{16\%}}\\
& Very        & \progressbar{36.667}{46.667}{13.333}{3.333}{0.000}{\textbf{83\%}}{\textbf{3\%}}\\
& Extremely   & \progressbar{35.484}{19.355}{35.484}{3.226}{6.452}{\textbf{55\%}}{\textbf{10\%}}\\

\bottomrule
\end{tabular}

\vspace{0.6em}
\likertlegend
\end{table*}

\begin{table*}[!htbp]
\centering
\small
\renewcommand{\arraystretch}{1.2}
\setlength{\tabcolsep}{7pt}

\caption{Contd. Likert distributions for preferences for project characteristics (OSS Practitioners). Percentages in bold show “Very/Extremely” (left) and “Not at all/Slightly” (right).}
\label{tab:projchars_op_continued_incentives}

\newcolumntype{L}[1]{>{\raggedright\arraybackslash}p{#1}}
\newcolumntype{C}[1]{>{\centering\arraybackslash}p{#1}}

\begin{tabular}{L{0.15\textwidth} L{0.15\textwidth} | C{0.60\textwidth}}
\toprule
\textbf{Motivation} & \textbf{Motivation Intensity} & \textbf{Distribution}\\
\midrule

\multirow{12}{*}{\textbf{Incentives}}

&  & \textit{A. Having Own Webpage}\\[-8pt]
& Not at all  & \progressbar{3.774}{13.208}{45.283}{24.528}{13.208}{\textbf{17\%}}{\textbf{38\%}}\\
& Slightly    & \progressbar{13.043}{21.739}{26.087}{34.783}{4.348}{\textbf{35\%}}{\textbf{39\%}}\\
& Moderately  & \progressbar{14.286}{33.333}{23.810}{23.810}{4.762}{\textbf{48\%}}{\textbf{29\%}}\\
& Very        & \progressbar{18.182}{36.364}{36.364}{9.091}{0.000}{\textbf{55\%}}{\textbf{9\%}}\\
& Extremely   & \progressbar{26.667}{13.333}{40.000}{20.000}{0.000}{\textbf{40\%}}{\textbf{20\%}}\\

\cmidrule(lr){2-3}

&  & \textit{B. Multilingual Documentation}\\[-2pt]
& Not at all  & \progressbar{9.434}{7.547}{13.208}{16.981}{52.830}{\textbf{17\%}}{\textbf{70\%}}\\
& Slightly    & \progressbar{4.348}{17.391}{8.696}{34.783}{34.783}{\textbf{22\%}}{\textbf{70\%}}\\
& Moderately  & \progressbar{9.524}{14.286}{28.571}{23.810}{23.810}{\textbf{24\%}}{\textbf{48\%}}\\
& Very        & \progressbar{27.273}{9.091}{9.091}{18.182}{36.364}{\textbf{36\%}}{\textbf{55\%}}\\
& Extremely   & \progressbar{13.333}{13.333}{46.667}{13.333}{13.333}{\textbf{27\%}}{\textbf{27\%}}\\

\midrule
\multirow{18}{*}{\textbf{Networking}}

&  & \textit{A. Having Own Webpage}\\[-8pt]
& Not at all  & \progressbar{0.000}{26.316}{21.053}{36.842}{15.789}{\textbf{26\%}}{\textbf{53\%}}\\
& Slightly    & \progressbar{8.000}{24.000}{28.000}{28.000}{12.000}{\textbf{32\%}}{\textbf{40\%}}\\
& Moderately  & \progressbar{11.111}{11.111}{50.000}{25.000}{2.778}{\textbf{22\%}}{\textbf{28\%}}\\
& Very        & \progressbar{13.793}{20.690}{37.931}{24.138}{3.448}{\textbf{34\%}}{\textbf{28\%}}\\
& Extremely   & \progressbar{28.571}{28.571}{35.714}{0.000}{7.143}{\textbf{57\%}}{\textbf{7\%}}\\

\cmidrule(lr){2-3}

&  & \textit{B. Multilingual Documentation}\\[-2pt]
& Not at all  & \progressbar{0.000}{15.789}{0.000}{21.053}{63.158}{\textbf{16\%}}{\textbf{84\%}}\\
& Slightly    & \progressbar{12.000}{12.000}{12.000}{16.000}{48.000}{\textbf{24\%}}{\textbf{64\%}}\\
& Moderately  & \progressbar{2.778}{5.556}{30.556}{27.778}{33.333}{\textbf{8\%}}{\textbf{61\%}}\\
& Very        & \progressbar{6.897}{17.241}{20.690}{24.138}{31.034}{\textbf{24\%}}{\textbf{55\%}}\\
& Extremely   & \progressbar{50.000}{7.143}{21.429}{7.143}{14.286}{\textbf{57\%}}{\textbf{21\%}}\\

\cmidrule(lr){2-3}

&  & \textit{C. Having Source Code Comments}\\[-2pt]
& Not at all  & \progressbar{10.526}{36.842}{10.526}{26.316}{15.789}{\textbf{42\%}}{\textbf{47\%}}\\
& Slightly    & \progressbar{12.000}{32.000}{32.000}{16.000}{8.000}{\textbf{44\%}}{\textbf{24\%}}\\
& Moderately  & \progressbar{25.000}{36.111}{33.333}{5.556}{0.000}{\textbf{61\%}}{\textbf{6\%}}\\
& Very        & \progressbar{27.586}{31.034}{37.931}{3.448}{0.000}{\textbf{59\%}}{\textbf{3\%}}\\
& Extremely   & \progressbar{50.000}{35.714}{7.143}{7.143}{0.000}{\textbf{86\%}}{\textbf{7\%}}\\

\bottomrule
\end{tabular}

\vspace{0.6em}
\likertlegend
\end{table*}

\begin{table*}[!htbp]
\centering
\small
\renewcommand{\arraystretch}{1.2}
\setlength{\tabcolsep}{7pt}

\caption{Contd. Likert distributions for preference for project characteristics (OSS Practitioners). Percentages in bold show “Very/Extremely” (left) and “Not at all/Slightly” (right).}
\label{tab:projchars_op_continued_enjoyment_helping}

\newcolumntype{L}[1]{>{\raggedright\arraybackslash}p{#1}}
\newcolumntype{C}[1]{>{\centering\arraybackslash}p{#1}}

\begin{tabular}{L{0.15\textwidth} L{0.15\textwidth} | C{0.60\textwidth}}
\toprule
\textbf{Motivation} & \textbf{Motivation Intensity} & \textbf{Distribution}\\
\midrule

\multirow{6}{*}{\textbf{Enjoyment}}
&  & \textit{A. Multilingual Documentation}\\[-2pt]
& Not at all  & \progressbar{0.000}{0.000}{50.000}{0.000}{50.000}{\textbf{0\%}}{\textbf{50\%}}\\
& Slightly    & \progressbar{0.000}{5.882}{23.529}{35.294}{35.294}{\textbf{6\%}}{\textbf{71\%}}\\
& Moderately  & \progressbar{2.941}{11.765}{20.588}{23.529}{41.176}{\textbf{15\%}}{\textbf{65\%}}\\
& Very        & \progressbar{13.725}{17.647}{17.647}{17.647}{33.333}{\textbf{31\%}}{\textbf{51\%}}\\
& Extremely   & \progressbar{26.316}{0.000}{10.526}{15.789}{47.368}{\textbf{26\%}}{\textbf{63\%}}\\

\midrule

\multirow{6}{*}{\textbf{Helping}}
&  & \textit{A. Multilingual Documentation}\\[-2pt]
& Not at all  & \progressbar{0.000}{0.000}{0.000}{0.000}{100.000}{\textbf{0\%}}{\textbf{100\%}}\\
& Slightly    & \progressbar{0.000}{6.250}{31.250}{43.750}{18.750}{\textbf{6\%}}{\textbf{62\%}}\\
& Moderately  & \progressbar{8.108}{5.405}{27.027}{27.027}{32.432}{\textbf{14\%}}{\textbf{59\%}}\\
& Very        & \progressbar{10.417}{20.833}{14.583}{14.583}{39.583}{\textbf{31\%}}{\textbf{54\%}}\\
& Extremely   & \progressbar{23.810}{4.762}{4.762}{9.524}{57.143}{\textbf{29\%}}{\textbf{67\%}}\\

\bottomrule
\end{tabular}

\vspace{0.6em}
\likertlegend
\end{table*}

\begin{table}[htbp]
\caption{OSS Practitioners' motivations and project characteristics statistics}
\label{tab:pract-pc-stat}
\centering
\begin{tabular}{@{}llllllll@{}}
\toprule
Motivation & Project Characteristics & H & df & p & k & $\epsilon^2$ & Effect \\ \midrule
Learning & Having clear guidelines & 18.183 & 4 & 0.001 & 5 & 0.120 & Medium \\
Reputation & Having Source code comments & 13.801 & 4 & 0.008 & 5 & 0.083 & Medium \\
Career & Having own web page & 10.316 & 4 & 0.035 & 5 & 0.054 & Small \\
 & Multilingual Documentation & 13.2 & 4 & 0.01 & 5 & 0.079 & Medium \\
 & Having Source code comments & 12.671 & 4 & 0.013 & 5 & 0.073 & Medium \\
Incentives & Having own web page & 13.367 & 4 & 0.01 & 5 & 0.079 & Medium \\
 & Multilingual Documentation & 11.27 & 4 & 0.024 & 5 & 0.062 & Medium \\
Networking & Having own web page & 11.485 & 4 & 0.022 & 5 & 0.063 & Medium \\
 & Multilingual Documentation & 16.806 & 4 & 0.002 & 5 & 0.109 & Medium \\
 & Having Source code comments & 19.283 & 4 & 0.001 & 5 & 0.129 & Medium \\
Enjoyment & Multilingual Documentation & 11.081 & 4 & 0.026 & 5 & 0.060 & Medium \\
Helping & Multilingual Documentation & 18.547 & 4 & 0.001 & 5 & 0.123 & Medium \\ 
\bottomrule
\end{tabular}
\end{table}

\begin{tcolorbox}[arc=1mm,width=\linewidth,
                  top=1mm,left=1mm,right=1mm,bottom=1mm,
                  boxrule=0.75pt,colback=gray!5,colframe=black!50]
\textbf{RQ 2 Summary:} \textcolor{black}{Motivations showed statistically significant correlations with project characteristic preferences.} Both \texttt{newcomers to OSS} and \texttt{OSS practitioners} consistently favored new projects in active development, small contributor groups, and clear contribution guidelines. Having source code comments was rated highly across all motivations, while multilingual documentation and follower counts were generally less important.
\end{tcolorbox}

\subsection {\textbf{Correlations between demographic factors and project characteristics (RQ3)}} 
\begin{figure*}
    \centering
    \includegraphics[width=\textwidth]{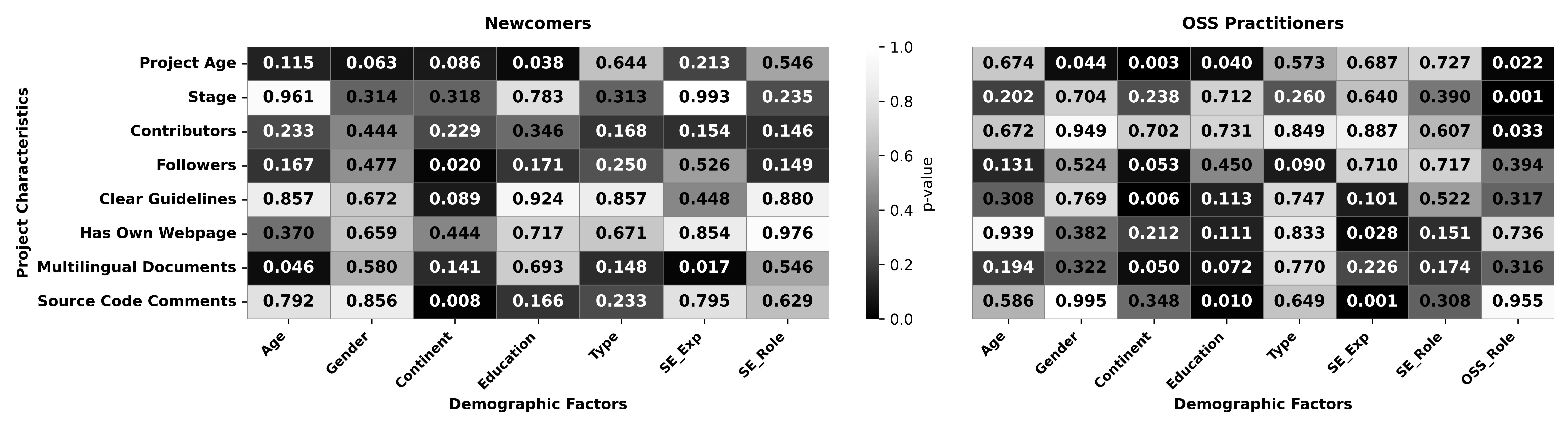}
    \caption{Demographic Factors Vs. Project Characteristics Likelihood Ratio Analysis \textit{p-values} Heat Maps}
    \label{correl_heatmap}
\end{figure*}

To explore the statistically significant differences between demographic factors and project characteristics, we performed the \textit{Kruskal-Wallis test} \citep{mckight2010kruskal} followed by Post-Hoc Bonferroni correction. All findings presented in this paper were statistically significant, with Bonferroni adjusted p-values less than 0.05. \textcolor{black}{Kruskal-Wallis test's p-value heat maps for Newcomers' and \texttt{OSS practitioners}' demographics and project characteristic preferences are presented in Figure \ref{correl_heatmap}.}

\subsubsection{\texttt{Newcomers}' correlations (\textbf{n=85})}

\texttt{newcomers to OSS} aged 18–24 were 26\% more likely to prefer multilingual documentation than those aged 35 and above (P = 0.004). Participants with postgraduate education favored new projects 36\% more than undergraduates (P = 0.029). Similarly, novices to the SE industry showed a 12\% higher preference for multilingual documentation compared to juniors (P = 0.018). In terms of preferences for followers, Participants from Africa were more likely to prefer small projects, with preferences 67\% higher than Oceania (P = 0.041) and 56\% higher than Europe (P = 0.026). Similarly, participants from Asia showed a 33\% stronger preference for small projects compared to Europe (P = 0.048) \textcolor{black}{(See Table \ref{tab:multilang_age_seexp_nc} for a detailed breakdown of the project characteristics preferences distributions).}

\textcolor{black}{
The effect size analysis in Table \ref{nc-demo-rq3} further indicated varying levels of practical significance across the identified correlations. The strongest practical effect was observed between Region and preferences for Followers ($\epsilon^2 = 0.164$), suggesting that regional differences had a substantial effect on newcomers’ preferences regarding project community size. The remaining statistically significant relationships demonstrated medium effect sizes.
}

\begin{table*}[!htbp]
\centering
\small
\renewcommand{\arraystretch}{1.15}
\setlength{\tabcolsep}{7pt}

\caption{Likert distributions for preferences towards Project Characteristics based on Demographics (Newcomers). Percentages in bold indicate: for 5-point Likert items, they correspond to “Very/Extremely” (left) and “Not at all/Slightly” (right), while for other Likert scales they represent the left-most and right-most response options.}
\label{tab:multilang_age_seexp_nc}

\newcolumntype{L}[1]{>{\raggedright\arraybackslash}p{#1}}
\newcolumntype{C}[1]{>{\centering\arraybackslash}p{#1}}

\begin{tabular}{L{0.2\textwidth} L{0.18\textwidth} | C{0.56\textwidth}}
\toprule
\textbf{Demographic Factor} & \textbf{Group} & \textbf{Distribution}\\
\midrule

\multirow{4}{*}{\textbf{Age}}
&  & \textit{Preferences for Multilingual Documentation}\\[-2pt]
& 18--24     & \progressbar{9.524}{23.810}{23.810}{19.048}{23.810}{\textbf{33\%}}{\textbf{43\%}}\\
& 25--34     & \progressbar{6.000}{8.000}{24.000}{24.000}{38.000}{\textbf{14\%}}{\textbf{62\%}}\\
& 35--above  & \progressbar{0.000}{7.143}{0.000}{21.429}{71.429}{\textbf{7\%}}{\textbf{93\%}}\\

&  & \footnotesize
\legendbox{Extremely}\, Extremely \hspace{0.9em}
\legendbox{Very}\, Very \hspace{0.9em}
\legendbox{Moderately}\, Moderately \hspace{0.9em}
\legendbox{Slightly}\, Slightly \hspace{0.9em}
\legendbox{NotAtAll}\, Not at all
\\

\midrule

\multirow{4}{*}{\textbf{SE Experience}}
&  & \textit{Preferences for Multilingual Documentation}\\[-2pt]
& Newcomer & \progressbar{22.222}{11.111}{33.333}{11.111}{22.222}{\textbf{33\%}}{\textbf{33\%}}\\
& Junior   & \progressbar{4.762}{16.667}{23.810}{30.952}{23.810}{\textbf{21\%}}{\textbf{55\%}}\\
& Senior   & \progressbar{2.941}{5.882}{11.765}{14.706}{64.706}{\textbf{9\%}}{\textbf{79\%}}\\

&  & \footnotesize
\legendbox{Extremely}\, Extremely \hspace{0.9em}
\legendbox{Very}\, Very \hspace{0.9em}
\legendbox{Moderately}\, Moderately \hspace{0.9em}
\legendbox{Slightly}\, Slightly \hspace{0.9em}
\legendbox{NotAtAll}\, Not at all
\\
\midrule

\multirow{4}{*}{\textbf{Education}}
&  & \textit{Preferences for Project Age}\\[-2pt]
& Basic     & \progressbarThree{50.000}{0.000}{50.000}{\textbf{50\%}}{\textbf{50\%}}\\
& Undergrad & \progressbarThree{48.148}{5.556}{46.296}{\textbf{48\%}}{\textbf{46\%}}\\
& Postgrad  & \progressbarThree{84.211}{0.000}{15.789}{\textbf{84\%}}{\textbf{16\%}}\\

&  & \footnotesize
\legendboxA{PBNew}\, New \hspace{1.0em}
\legendboxA{PBMature}\, Mature \hspace{1.0em}
\legendboxA{PBNeutral}\, Doesn't matter
\\

\midrule

\multirow{6}{*}{\textbf{Region}}
&  & \textit{Preferences for Followers}\\[-2pt]
& Africa  & \progressbarFour{66.667}{16.667}{0.000}{16.667}{\textbf{67\%}}{\textbf{17\%}}\\
& America & \progressbarFour{35.000}{5.000}{0.000}{60.000}{\textbf{35\%}}{\textbf{60\%}}\\
& Asia    & \progressbarFour{44.444}{5.556}{11.111}{38.889}{\textbf{44\%}}{\textbf{39\%}}\\
& Europe  & \progressbarFour{11.111}{5.556}{5.556}{77.778}{\textbf{11\%}}{\textbf{78\%}}\\
& Oceania & \progressbarFour{0.000}{0.000}{0.000}{100.000}{\textbf{0\%}}{\textbf{100\%}}\\

&  & \footnotesize
\legendboxA{FSmall}\, Small \hspace{1.0em}
\legendboxA{FMedium}\, Medium \hspace{1.0em}
\legendboxA{FLarge}\, Large \hspace{1.0em}
\legendboxA{FNeutral}\, Doesn't matter
\\

\bottomrule
\end{tabular}
\end{table*}

\begin{table}[htbp]
\caption{Newcomer’s demographics and project characteristics statistics}
\label{nc-demo-rq3}
\centering
\begin{tabular}{@{}llllllll@{}}
\toprule
Demographic & Project Characteristics & H & df & p & k & $\epsilon^2$ & Effect \\ \midrule
Age & Multilingual Documentation & 10.086 & 2 & 0.006 & 3 & 0.099 & Medium \\
SE experience & Multilingual Documentation & 13.054 & 2 & 0.001 & 3 & 0.135 & Medium \\
Education & Project Age & 7.093 & 2 & 0.029 & 3 & 0.062 & Medium \\
Region & Preference for Followers & 17.101 & 4 & 0.002 & 5 & 0.164 & Large \\ 
\bottomrule
\end{tabular}
\end{table}

\subsubsection{\texttt{OSS practitioners}' correlations (\textbf{n=123})}

For the project age, 68\% more females preferred new projects compared to the ``other'' gender category (p-value 0.033). Furthermore, 34\% more participants from Africa preferred new projects compared to those from Europe (p-value 0.049). The inclination to have clear guidelines varies by region, with 39\% more respondents from Africa preferring it compared to Europe (p-value 0.004). Moreover, Africa demonstrated a significantly greater inclination towards offering multilingual documentation, with a 24\% preference compared to America (p-value 0.004).

Juniors and Seniors demonstrated a significantly greater inclination towards having their own webpage with 35\% and 31\% higher preferences compared to newcomers to SE (p-values 0.021). Similarly, Juniors and Seniors demonstrated a significantly greater preference for having source code comments with 69\% and 53\% higher preferences compared to newcomers to SE (p-values 0.002, 0.008) \textcolor{black}{(See Table \ref{tab:projchars_op_table10_style} for a detailed breakdown of project characteristics preferences distributions).}

\textcolor{black}{
The effect size analysis in Table \ref{pc-demo-rq3} further indicated varying levels of practical significance across the identified correlations. The strongest practical effects were observed between Region and Having Clear Guidelines ($\epsilon^2 = 0.099$), and SE Experience and Having Source Code Comments ($\epsilon^2 = 0.084$), both demonstrating medium effect sizes. Region and Having Multilingual Documentation also showed a medium practical effect ($\epsilon^2 = 0.076$). In contrast, Region and Project Age ($\epsilon^2 = 0.013$), Gender and Project Age ($\epsilon^2 = 0.048$) and SE Experience and Having Own Web Page ($\epsilon^2 = 0.047$) demonstrated smaller practical effects, suggesting comparatively weaker effects on OSS practitioners' project characteristic preferences.
}

\begin{table*}[]
\centering
\small
\renewcommand{\arraystretch}{1.15}
\setlength{\tabcolsep}{7pt}

\caption{Likert distributions for preference towards Project Characteristics based on Demographics (OSS Practitioners). Percentages in bold indicate: for 5-point Likert items, they correspond to “Very/Extremely” (left) and “Not at all/Slightly” (right), while for other Likert scales they represent the left-most and right-most response options.}
\label{tab:projchars_op_table10_style}

\newcolumntype{L}[1]{>{\raggedright\arraybackslash}p{#1}}
\newcolumntype{C}[1]{>{\centering\arraybackslash}p{#1}}

\begin{tabular}{L{0.15\textwidth} L{0.18\textwidth} | C{0.56\textwidth}}
\toprule
\textbf{Factor} & \textbf{Group} & \textbf{Distribution}\\
\midrule

\multirow{4}{*}{\textbf{Gender}}
&  & \textit{Preferences for Project Age}\\[-2pt]
& Female & \progressbarThree{68}{12}{20}{\textbf{68\%}}{\textbf{20\%}}\\
& Male   & \progressbarThree{49.474}{10.526}{40}{\textbf{49\%}}{\textbf{40\%}}\\
& Other  & \progressbarThree{0}{0}{100}{\textbf{0\%}}{\textbf{100\%}}\\

&  & \footnotesize
\legendboxA{PBNew}\, New \hspace{1.0em}
\legendboxA{PBMature}\, Mature \hspace{1.0em}
\legendboxA{PBNeutral}\, Doesn't matter
\\

\midrule

\multirow{6}{*}{\textbf{Region}}
&  & \textit{Preferences for Project Age}\\[-2pt]
& Africa  & \progressbarThree{81.250}{0}{18.750}{\textbf{81\%}}{\textbf{19\%}}\\
& America & \progressbarThree{56.667}{0}{43.333}{\textbf{57\%}}{\textbf{43\%}}\\
& Asia    & \progressbarThree{44.000}{24.000}{32.000}{\textbf{44\%}}{\textbf{32\%}}\\
& Europe  & \progressbarThree{46.667}{8.889}{44.444}{\textbf{47\%}}{\textbf{44\%}}\\
& Oceania & \progressbarThree{28.571}{42.857}{28.571}{\textbf{29\%}}{\textbf{29\%}}\\

&  & \footnotesize
\legendboxA{PBNew}\, New \hspace{1.0em}
\legendboxA{PBMature}\, Mature \hspace{1.0em}
\legendboxA{PBNeutral}\, Doesn't matter
\\

\midrule

\multirow{6}{*}{\textbf{Region}}
&  & \textit{Preferences for Clear Guidelines}\\[-2pt]
& Africa  & \progressbar{56.250}{31.250}{12.500}{0}{0}{\textbf{88\%}}{\textbf{0\%}}\\
& America & \progressbar{16.667}{46.667}{30.000}{0}{6.667}{\textbf{63\%}}{\textbf{7\%}}\\
& Asia    & \progressbar{48.000}{20.000}{28.000}{4.000}{0}{\textbf{68\%}}{\textbf{4\%}}\\
& Europe  & \progressbar{13.333}{35.556}{37.778}{13.333}{0}{\textbf{49\%}}{\textbf{13\%}}\\
& Oceania & \progressbar{28.571}{28.571}{42.857}{0}{0}{\textbf{57\%}}{\textbf{0\%}}\\

&  & \footnotesize
\legendbox{Extremely}\, Extremely \hspace{0.9em}
\legendbox{Very}\, Very \hspace{0.9em}
\legendbox{Moderately}\, Moderately \hspace{0.9em}
\legendbox{Slightly}\, Slightly \hspace{0.9em}
\legendbox{NotAtAll}\, Not at all
\\

\midrule

\multirow{6}{*}{\textbf{Region}}
&  & \textit{Preferences for Multilingual Documentation}\\[-2pt]
& Africa  & \progressbar{25.000}{18.750}{31.250}{12.500}{12.500}{\textbf{44\%}}{\textbf{25\%}}\\
& America & \progressbar{6.667}{13.333}{3.333}{20.000}{56.667}{\textbf{20\%}}{\textbf{77\%}}\\
& Asia    & \progressbar{8.000}{4.000}{32.000}{24.000}{32.000}{\textbf{12\%}}{\textbf{56\%}}\\
& Europe  & \progressbar{11.111}{11.111}{20.000}{22.222}{35.556}{\textbf{22\%}}{\textbf{58\%}}\\
& Oceania & \progressbar{0}{14.286}{0}{28.571}{57.143}{\textbf{14\%}}{\textbf{86\%}}\\

&  & \footnotesize
\legendbox{Extremely}\, Extremely \hspace{0.9em}
\legendbox{Very}\, Very \hspace{0.9em}
\legendbox{Moderately}\, Moderately \hspace{0.9em}
\legendbox{Slightly}\, Slightly \hspace{0.9em}
\legendbox{NotAtAll}\, Not at all
\\

\midrule

\multirow{4}{*}{\textbf{SE Experience}}
&  & \textit{Preferences for Own Webpage}\\[-2pt]
& Newcomer & \progressbar{0}{0}{0}{80.000}{20.000}{\textbf{0\%}}{\textbf{100\%}}\\
& Junior   & \progressbar{11.538}{23.077}{38.462}{15.385}{11.538}{\textbf{35\%}}{\textbf{27\%}}\\
& Senior   & \progressbar{12.121}{19.697}{37.879}{27.273}{3.030}{\textbf{32\%}}{\textbf{6\%}}\\

&  & \footnotesize
\legendbox{Extremely}\, Extremely \hspace{0.9em}
\legendbox{Very}\, Very \hspace{0.9em}
\legendbox{Moderately}\, Moderately \hspace{0.9em}
\legendbox{Slightly}\, Slightly \hspace{0.9em}
\legendbox{NotAtAll}\, Not at all
\\

\midrule

\multirow{4}{*}{\textbf{SE Experience}}
&  & \textit{Preferences for Source Code Comments}\\[-2pt]
& Newcomer & \progressbar{0}{0}{20.000}{40.000}{40.000}{\textbf{0\%}}{\textbf{80\%}}\\
& Junior   & \progressbar{28.846}{40.385}{17.308}{7.692}{5.769}{\textbf{69\%}}{\textbf{13\%}}\\
& Senior   & \progressbar{21.212}{31.818}{36.364}{10.606}{0}{\textbf{53\%}}{\textbf{11\%}}\\

&  & \footnotesize
\legendbox{Extremely}\, Extremely \hspace{0.9em}
\legendbox{Very}\, Very \hspace{0.9em}
\legendbox{Moderately}\, Moderately \hspace{0.9em}
\legendbox{Slightly}\, Slightly \hspace{0.9em}
\legendbox{NotAtAll}\, Not at all
\\

\bottomrule
\end{tabular}
\end{table*}

\begin{table}[htbp]
\caption{OSS Practitioners' demographics and project characteristics statistics}
\label{pc-demo-rq3}
\centering
\begin{tabular}{@{}llllllll@{}}
\toprule
Demographic & Project Characteristics & H & df & p & k & $\epsilon^2$ & Effect \\ \midrule
Gender & Project Age & 7.734 & 2 & 0.021 & 3 & 0.048 & Small \\
Region & Project Age & 5.476 & 4 & 0.242 & 5 & 0.013 & Small \\
 & Having clear guidelines & 15.714 & 4 & 0.003 & 5 & 0.099 & Medium \\
 & Multilingual Documentation & 12.947 & 4 & 0.012 & 5 & 0.076 & Medium \\
SE experience & Having own web page & 7.593 & 2 & 0.022 & 3 & 0.047 & Small \\
 & Having Source code comments & 12.061 & 2 & 0.02 & 3 & 0.084 & Medium \\ 
 \bottomrule
\end{tabular}
\end{table}

\begin{tcolorbox}[arc=1mm,width=\linewidth,
                  top=1mm,left=1mm,right=1mm,bottom=1mm,
                  boxrule=0.75pt,colback=gray!5,colframe=black!50]
\textbf{RQ 3 Summary:} Clear demographic patterns emerged in project preferences. \texttt{Newcomers'} choices varied by age, education, SE experience, and region, with younger and less experienced participants emphasizing multilingual documentation and African/Asian respondents preferring smaller projects. Among \texttt{OSS practitioners}, gender, region, and SE role influenced preferences for project age, having clear guidelines, multilingual documentation, having their own webpages, and source code comments.
\end{tcolorbox}

\subsection{\textbf{ Practitioners' views on improving recommendations (RQ4)}}
The analysis of practitioners' views on improving recommendation systems to better align with their motivations has identified four key themes, which are described below. Table \ref{code table} shows the key themes and relevant code examples identified during thematic analysis.

\begin{table}[t]
\caption{Themes and code examples identified during thematic analysis}
\label{code_table}
\small
\renewcommand{\arraystretch}{1.1}
\begin{tabularx}{\linewidth}{@{}p{3.5cm}X@{}}
\toprule
\textbf{Themes} & \textbf{Code Examples} \\
\midrule

Personal Fit \& Growth Opportunities (68)
&
recommendations for new technologies, develop skills in high-demand areas,
user current skill level with options to advance
\\
\midrule

Project Characteristics (42)
&
high-profile, complexity, well-documented code, community reputation,
real-world impact, time-bound challenges
\\
\midrule

Interactive \& Dynamically Evolving Recommendation Process (25)
&
questionnaire to personalize, feedback loop, learn more and adapt,
intermittent check-ins, connect to social networks
\\
\midrule

Collaboration \& Team Dynamics (21)
&
projects with large communities, team sizes, networking opportunities,
reputable contributors, kind, patient, encouraging
\\

\bottomrule
\end{tabularx}
\end{table}

    \textit{\textbf{Personal Fit \& Growth Opportunities}}. This theme emerged as the most pronounced, with participants emphasizing the need to recommend projects that align with their current experience and skills, as well as projects involving new technologies. 

\begin{quoting}
   ``\textit{I think it should also select recommendations for new technologies that I have not yet met, so that I can be up to date and ready for future challenges.}'' (Developer)
\end{quoting}

     Moreover, they mentioned that the recommendations should align with the domains they are interested in and offer opportunities for learning, achieving personal goals and advancing their careers.

     \begin{quoting}
   ``\textit{Most importantly, recommendations should ideally align with my interests (either the problem domain and/or technologies used). Sheer popularity of the project is less important to me as the main career benefit comes in the form of experience and interesting stories for interviews that demonstrate my collaborative and technical capabilities.}'' (Developer)
\end{quoting}

   \textit{\textbf{Project Characteristics}}. The participants highlighted the key project characteristics they consider when selecting an OSS project to contribute to. The most frequently mentioned characteristic was the project's reputation, particularly when it is backed by well-recognized companies, which adds credibility to the project.
   
    \begin{quoting}
   ``\textit{High-profile projects backed by recognized companies would fully align with my motivations}'' (Developer)
\end{quoting}
   
   Further, they value the real-world impact of the project, considering factors such as clear guidelines and documentation, tech stack, project state, complexity, sponsorships and financial rewards, time-bound tasks, test coverage and frequency of updates.

    \begin{quoting}
   ``\textit{My motivation is to make people’s lives better by making apps that make their tech approach easier, so I would prefer projects with specific guidelines or concrete ideas.}'' (Developer)
\end{quoting}

\begin{quoting}
   ``\textit{Adding filters in addition to programming language, like state of the project so it is known if there's code or IU missing or just testing for bugs}'' (Developer)
\end{quoting}

\textit{\textbf{Interactive \& Dynamically Evolving Recommendation Process}}. The participants valued the ability to interact with the recommendation process by providing preferences and feedback. They suggested introducing filters, answering a questionnaire, or providing instructions using simple text prompts as the interaction methods.

\begin{quoting}
   ``\textit{Project recommendation systems can be improved by allowing filtering and categorization, for example through AI, so that a list of project that fit my criteria can be invoked with a simple one line prompt.}'' (Developer)
\end{quoting}

Additionally, they suggested transferring knowledge from social media platforms like LinkedIn. Further, they emphasized that it is important to update and adapt recommendations regularly.

\begin{quoting}
   ``\textit{Link the system into linkedin and interests}'' (AI/ML Engineer)
\end{quoting}

\begin{quoting}
   ``\textit{Recommendation systems can do intermittent check-ins to collect data on the user’s current motivations, which can evolve over time. }'' (Developer)
\end{quoting}

\textit{\textbf{Collaboration \& Team Dynamics }}. Collaboration, networking opportunities, and varied team dynamics were appreciated in the responses. Participants emphasized the importance of having larger communities or team sizes, reputable contributors, and collaborators they have previously worked with. 

\begin{quoting}
   ``\textit{Test me, then recommend me projects I could collaborate on and would help me improve to collaborate on more.}'' (Developer)
\end{quoting}

Further, they highlighted the value of active members who are kind, patient, and encouraging within the community.

\begin{quoting}
   ``\textit{I would want to know which communities are known for being kind, patient, and encouraging. I'm not as concerned with prestige. Too much ego will kill an OSS project. I'd much rather work with humble contributors and not be recognized.}'' (Bug reporter)
\end{quoting}

\begin{tcolorbox}[arc=1mm,width=1.0\columnwidth,
                  top=1mm,left=1mm,  right=1mm, bottom=1mm,
                  boxrule=.75pt]
\textbf{RQ 4 Summary:} Practitioners highlighted four main themes for improving recommendation systems. Personal Fit \& Growth Opportunities with projects that align with their current skills and involve new technologies, importance of Project Characteristics such as the project’s reputation, complexity and the real-world impact of the project. Moreover, they highlighted the importance of Interactive \& Dynamically Evolving Recommendation Systems that allow them to interact with the system, and Collaboration \& Team Dynamics, which support networking opportunities.
\end{tcolorbox}

\section{Discussion} \label{discussion}

 \textcolor{black}{This section presents potential design directions for future OSS recommendation systems, along with implications for researchers and practitioners, based on contributors’ stated preferences and experiences.}

\textbf{\textit{OSS project owners and companies are suggested to structure contribution pathways to align with motivations and influencing factors, \textcolor{black}{potentially supporting contributor engagement and onboarding experiences.
}}}

\textcolor{black}{Our results (RQ1) show statistically significant correlations between demographic attributes and specific motivations, while RQ2 shows that contributors with different motivations reported different preferences for project characteristics.} A healthy OSS community brings together a diverse range of contributors and sustains itself by retaining them \citep{guizani2025community}. Hence, OSS project owners \textcolor{black}{may consider how projects can better reflect the needs of contributors from diverse backgrounds to build inclusive and sustainable communities.} 

By tagging tasks with labels like \textit{Learning Opportunity},\textit{ Community Aid}, or \textit{Creative Exploration}, \textcolor{black}{projects may make motivation relevant opportunities more visible to contributors with different demographically influenced preferences (RQ1)}. Embedding project-level cues like \textit{Development Stage} or \textit{``Multilingual Documentation''} into \textcolor{black}{these tags could help contributors identify tasks that appear aligned with their motivations and personal goals.} Moreover, RQ3's results highlight correlations between demographics and project characteristics, such as \texttt{newcomers to OSS} from Africa and Asia preferring projects with a small number of followers. OSS project owners \textcolor{black}{may use these insights as preliminary guidance when defining and advertising their projects.} For example, if a project targets contributors from specific experience levels, regions or education levels, maintainers can highlight these characteristics in the project description. Additionally, they can label repositories with demographic-aligned features (e.g., ``Beginner-Friendly: Small Community").

Additionally, projects can introduce recognition mechanisms, such as badges or micro-credentials linked to motivation-aligned contributions. For example, contributors could earn badges like \textit{Skill Builder} for resolving complex issues or \textit{Community Helper} for addressing help-wanted issues or support tickets. These recognition signals \textcolor{black}{could} be displayed in contributor profiles and utilized by \textcolor{black}{future} recommendation systems to match users with repositories offering unearned, motivation-aligned opportunities. Prior research on attracting and retaining OSS contributors through a maintainer dashboard found that introducing a “Rising Contributor” badge helped active \texttt{newcomers to OSS} feel recognized and gain credibility that encouraged them to stay engaged and continue contributing \citep{guizani2022attracting}.

Motivation tagging not only facilitates better discoverability of tasks but also supports targeted and equitable onboarding \citep{xiao2023personalized}. However, implementing such a system requires careful attention to other project characteristics, such as the governance model. Prior research has highlighted that when introducing recognition badges, it’s important to consider how contributors are compensated as OSS increasingly includes both paid and volunteer contributors working side by side \citep{guizani2022attracting}. While these badges might not significantly impact the motivation of paid contributors, they can hold considerable value for volunteers \citep{guizani2022attracting}.
Moreover, projects that display too many badges can lead to negative perceptions, making the project appear cluttered or less credible. It can backfire by overwhelming and distracting visitors from the project's important information \citep{trockman2018adding}.
   
\textbf{\textit{OSS recommendation system builders are suggested to design dynamically evolving, interactive recommendation systems along with comprehensive maintainers' dashboards}.}

The results (RQ4) indicate that practitioners prefer interactive and dynamically evolving recommendations. This need for personalization is \textcolor{black}{consistent with} our findings in RQ3. Since we found significant variation in project characteristic preferences by demographic traits, \textcolor{black}{future systems may benefit from} allowing customization along these demographic dimensions. Hence, OSS tool builders \textcolor{black}{could explore recommendation systems that allow users to actively shape recommendations through interactive features such as filters, questionnaires, or simple text prompts.} For example, Large language models (LLMs) can be used as conversational recommendation systems to enhance user engagement and personalization \citep{chen2025evaluating}. They can effectively process and understand natural language input while leveraging context from user queries \citep{chen2025evaluating, kim2025extracting}. Further, they can mitigate the cold-start problem (users who don’t have enough activity or historical data) by asking users relevant questions about preferences and analyzing contextual information. However, LLMs can be resource-intensive and often underperform in specialized domains without fine-tuning, as they are trained on broad datasets \citep{kim2025extracting}. Additionally, ethical considerations such as privacy and transparency may arise as processing user inputs, especially sensitive information, raises concerns about data privacy and security \citep{kim2025extracting}.

Further, the practitioners suggested (RQ4) transferring knowledge from other platforms like LinkedIn to enhance recommendation accuracy. This can address data sparsity by leveraging existing user preferences, interactions, and expertise \citep{song2022collaboration,talebzadeh2024evaluating}. This approach \textcolor{black}{may enable more personalized project matching, especially for new or less active contributors.} However, sharing user data across platforms poses potential privacy risks. Unauthorized access or data breaches during transmission can expose sensitive data \citep{wang2025federated}. Hence, strong security measures and regulatory compliance are essential. Moreover, ensuring seamless data exchange is technically challenging, as it requires compatibility across various platforms and data formats \citep{fu2024exploring}.

In addition, recent research highlights several complementary innovations that can enrich recommendation systems. Community monitoring dashboards, such as Community Tapestry \citep{guizani2025community}, dynamically track project health signals, including contributor turnover and diversity, allowing maintainers to detect declining newcomer retention and adjust outreach strategies accordingly. Climate Coach \citep{qiu2023climate} is another dashboard that helps OSS maintainers track community health by monitoring signals like responsiveness to issues, pushback in code reviews, and toxicity in discussions. It displays various dimensions of social aspects within the team, which can help maintainers improve their management. These signals suggest that, beyond recommending projects to contributors, building comprehensive dashboards for maintainers \textcolor{black}{may support OSS sustainability efforts}. We suggest extending these dashboards to monitor which contributor motivations a project supports, using the findings from our RQ2. It \textcolor{black}{may} help in attracting and retaining the right contributors.

\textbf{\textit{OSS contributors \textcolor{black}{may benefit from considering projects that offer} AI-powered tools support, gamified onboarding, and inclusive community practices}} 

Our findings in RQ4 suggest that, although personal fit and skill alignment are crucial considerations in selecting projects, practitioners should also look for opportunities that allow them to engage with new technologies and stay updated on industry trends. Engaging in projects that push technological boundaries can foster continuous learning and career advancement, ensuring long-term relevance in an evolving software landscape. Empirical evidence demonstrated that contributors using AI-powered tools such as GitHub Copilot experienced notable improvements in productivity and satisfaction, especially within the context of OSS \citep{song2024impact}. Moreover, having AI tool support may help motivate developers by offering practical solutions that align with their needs. For example, for those driven by Helping, these tools may reduce the workload
on contributors by automating repetitive tasks such as code reviews, bug detection, and security checks \citep{cihan2024automated, zhang2024fixing}. This may ultimately result in saving them time to help the community.

Further, the findings (RQ4) of this study highlight participants’ preference for communities that are not only technically active but also socially supportive, where kindness, patience, and encouragement are valued over prestige or visibility. Contributing to environments that prioritize constructive interactions and teamwork not only enhances individual experiences but also strengthens the overall health of OSS communities. Projects with higher psychological safety, where people can speak up without fear, retain contributors much longer \citep{sesari2025safe}. Moreover, studies have shown that bias or harsh treatment drives people away, whereas inclusive communities (which actively discourage harassment) attract diverse contributors and sustain growth \citep{li2021code, alebachew2025we}. Hence, before committing, practitioners should review the project's issue tracker and discussion forums for courteous, constructive interactions, look for a visible code of conduct or charter, and observe whether maintainers provide patient explanations and acknowledge contributions positively \citep{jamieson2024predicting,li2021code, alebachew2025we}.

For newcomers, gamified systems can provide an additional layer of support during onboarding. Recent studies have shown that quest-based platforms like OSSDoorway \citep{santos2025ossdoorway} have significantly improved contributors’ confidence and helped them navigate essential tasks, such as using GitHub and submitting pull requests. Hence, \texttt{newcomers to OSS} \textcolor{black}{may benefit from choosing projects that support such systems, as they may ease onboarding and build confidence early in the contribution journey}.

\textbf{\textit{Researchers are suggested to explore how contributors’ motivations evolve over time, vary across various OSS domains, and how correlations with influencing factors shift as contributors take on different roles over time.}} 

Our study offers a dual perspective on \texttt{newcomers to OSS} and experienced \texttt{OSS practitioners} separately, \textcolor{black}{showing how their reported motivations and preferences differ across groups}. \textcolor{black}{RQ1 showed that reported motivations vary across demographic groups, while RQ2 showed that different motivations are correlated with different stated preferences for project characteristics.} Further, RQ3 highlighted correlations between demographics and project characteristics, underscoring how different groups approach project selection in distinct ways.

Moreover, Motivations of OSS contributors evolve over time, often shifting from extrinsic to intrinsic ones \citep{gerosa2021shifting}. Hence, building on our findings (RQ1–RQ3), longitudinal studies could track individuals over time to capture how their motivations and project characteristic preferences shift in relation to demographic changes such as role transitions (e.g., from casual to core contributor) or experience gains. This would help validate whether the statistically significant patterns we observed persist, change, or reverse as contributors gain experience, ultimately advancing our understanding of contributor engagement dynamics.

Existing studies have examined the quality of communication influence on contributor retention \citep{wang2024community,jamieson2024predicting}. However, there is a limitation in comprehensively analyzing whether demographic-based personalization improves contributor retention and encourages long-term contributions. Future research should also explore how project-level practices either facilitate or hinder these motivational transitions. For instance, studies could evaluate whether explicit contribution pathways, or regular feedback loops support long-term sustained engagement. Moreover, our results (RQ1–RQ3) indicate several correlations with motivations, demographics and project characteristics. However, we did not explore these correlations across different project domains and motivations. Hence, examining how motivation varies across different OSS domains (e.g., Web \& Application Development projects vs AI \& Machine Learning projects) could provide tailored recommendation strategies. 

\section{Threats to Validity} \label{threats}
\textbf{\textit{Internal Validity:}} Several factors may have negatively influenced our data collection and analysis. Some responders may have lacked the experience and expertise necessary to participate in our study. To mitigate this, we implemented a screening test for participants before the survey began.
One more potential threat is recall bias \citep{tayeb2024investigating,gerosa2021shifting}, where participants might not accurately remember all the details \citep{tayeb2024investigating}. Another concern is social desirability bias, which occurs when individuals provide answers they think align with societal expectations rather than reflecting their true preferences or experiences. To mitigate these problems, we ensured that the data collection process was anonymous, encouraging participants to give more honest and accurate responses. Although we selected the most widely studied OSS motivations and project characteristics based on prior literature, our selection may not fully capture the complete range of contributors’ motivations and project characteristic preferences. Another potential threat to validity is the clarity and interpretation of survey questions, as ambiguous or unfamiliar terminology could have led to misinterpretations by participants. However, through our pilot study, we identified and addressed such issues based on the feedback received. Although we used a non-parametric test, the relatively small final sample size (208 responses) may have limited our ability to detect all significant correlations or relationships in the analysis.

\textbf{\textit{Construct Validity:}} A concern in our survey was the possibility of not including all the relevant options. This could potentially lead to underrepresentation or bias in our findings. To mitigate this, we included an “Other” option in relevant questions. However, the results show that most participants selected “Other” as their least preferred option. This suggests that our survey effectively captured the most commonly relevant options. Another potential threat is the subjectivity involved in interpreting responses during the thematic analysis process. To address this, \textcolor{black}{the first author initially coded the responses, while the second and third authors independently reviewed subsets of the coded data.} Any differences in their interpretations were resolved through discussions.

\textbf{\textit{External Validity:}} External validity threats arise from the limited generalizability of our findings, as our sample of 208 respondents may not fully represent the diverse OSS community. The over-representation of certain demographics or geographical regions within our sample could have influenced the results, limiting their applicability to the broader OSS contributor population. This highlights the necessity for future research to involve a more diverse and widespread population in order to \textcolor{black}{further evaluate} and expand upon our findings.
\textcolor{black}{Additionally, participants were recruited through both Prolific, where respondents received monetary compensation, and social media platforms, where participation was voluntary. Differences in recruitment context and compensation may have influenced response patterns, which could affect the generalizability of the findings.
}

\section{Conclusion \& Future Work } \label{conclusion}
This study provides a comprehensive analysis of the motivations, demographics and project characteristic preferences of both OSS \texttt{newcomers to OSS}and experienced practitioners. We revealed significant demographic \textcolor{black}{correlations}, with factors such as age, gender, and OSS role shaping motivations. Furthermore, project characteristics, including documentation quality, development stage, and project age, are strongly tied to specific motivations.  Additionally, practitioners expressed the need for dynamically evolving interactive recommendation systems that account for career growth, technical interests, and collaboration dynamics. \textcolor{black}{ By leveraging these insights, OSS communities can support more effective contributor onboarding, engagement, retention, and project-selection support practices within OSS communities. Additionally, they may inform the future design of recommendation systems and other support tools that better align with contributors' motivations}. 

Roles in OSS projects extend beyond coding-related and project-centric roles \citep{Trinkenreich2020HiddenFigures}. Community-centric roles such as advocates, strategists, mentors, license managers, and writers play a crucial part \citep{Trinkenreich2020HiddenFigures}. Hence, future research can build on our work by focusing specifically on these under-represented groups, exploring their motivations and how demographics and project characteristics influence their decisions to join a project. Further, future work can explore longitudinal studies to examine how motivations evolve over time and how motivations differ across various OSS domains. Additionally, research can investigate how emotions like passion, frustration, and burnout influence contributors’ motivations across varying project contexts.

\section{Declarations}

\subsection*{Funding}
This work was supported by the Deakin University Postgraduate Research Scholarship (DUPR-ROUND 0000018830).

\subsection*{Ethical Approval}

This study received ethics approval from the Deakin University Human Research Ethics Committee (Reference Number: 2024/HE000335).

\subsection*{Informed Consent}

Informed consent was obtained from all individual participants included in the study.

\subsection*{Author Contributions}

Shashiwadana Nirmani conducted data collection, data analysis, and manuscript writing. Hourieh Khalajzadeh and Mojtaba Shahin contributed to data analysis, manuscript review \& editing and supervision. Xiao Liu contributed to manuscript review \& editing and supervision.

\subsection*{Data Availability Statement}

The replication package is available at \href{https://doi.org/10.5281/zenodo.18764186}{https://doi.org/10.5281/zenodo.18764186}.

\subsection*{Conflict of Interest}

The authors declare that they have no conflict of interest.

\subsection*{Clinical Trial Number }
Not applicable.

\bibliography{ref}
\clearpage

\appendix

\section{Appendix - A} \label{Appendix-a}

\begin{figure*}[h] 
  \centering

\begin{surveybox}

\textbf{QUESTION 1:} Which of these programming languages have you worked with before?\\[0.25em]
\textbf{OPTIONS}: \texttt{C\# | C | C++ | Python | Javascript | Java | Ruby | PHP | Shell | Typescript | Other | I don't program}
\vspace{0.8em}

\textbf{QUESTION 2:} Which of these lesser-known programming languages have you worked with before?\\[0.25em]
\textbf{OPTIONS}: \texttt{Yod | Lore | Torg | Lprime | ThreeP | EMH | Holly | SHROUD | LTCdata | Kryten | None of the above}
\vspace{0.8em}

\textbf{Refer to the following code fragment to answer the next two questions.}\\[0.25em]
\begin{lstlisting}[language=Java]
main {
    print(func("hello world"))
}

String func(String in) {
    int x = len(in);
    String out = "";
    for (int i = x - 1; i >= 0; i--) {
        out.append(in[i]);
    }
    return out;
}
\end{lstlisting}

\textbf{QUESTION 3:} What is the parameter of the function?\\[0.25em]
\textbf{OPTIONS}: \texttt{String in | I don't know | int i = x-1; i >= 0; i- | Outputting a String | int x = len(in)}
\vspace{0.8em}

\textbf{QUESTION 4:} Please select the returned value of the pseudocode above.\\[0.25em]
\textbf{OPTIONS}: \texttt{hello world | hello world 10 | dlrow olleh | world hello | HELLO WORLD | I don't know | hello world hello world hello world hello world}

\end{surveybox}

\caption{Screening Questions}
  \label{fig:screening}
\end{figure*}

\clearpage

\section{Appendix -B}

\begin{table*}[h]
\centering
\footnotesize
\caption{Main survey questions. The full survey instrument including all response options for the single-choice questions (SCQ) and multiple-choice questions (MCQ) is provided in replication package \citep{anonymous2025motivation}.}
\label{tab:survey_questions}
\begin{tabularx}{\textwidth}{@{}p{0.06\textwidth} Y p{0.10\textwidth} p{0.22\textwidth}@{}}
\toprule
\textbf{Question No.} & \textbf{Survey Question (\emph{* indicates required})} & \textbf{Type} & \textbf{Notes} \\
\midrule

Q2  & To which gender identity do you most identify?* & \SCQ & -- \\
Q3  & Which continent are you originally from?* & \SCQ & Options include``Prefer not to say'' \\
Q4  & In which country do you currently reside?* & \SCQ & Options include ``Prefer not to say'' \\
Q5  & What is your highest level of education completed?* & \SCQ & -- \\
Q6  & What type of programmer are you?* & \SCQ & -- \\
Q7  & How many years of work experience do you have in the software industry?* & \SCQ & -- \\
Q8  & Given that you have worked in the software industry, what is your current role in the software industry?* & \SCQ & Shown if ``Never worked'' is not selected for Q7 \\
Q9  & You mentioned that your role is not listed. Please specify your current role.* & \FT & Shown if ``Other'' is selected for Q8 \\
Q10 & How many years of contribution experience do you have in OSS projects?* & \SCQ & -- \\
Q11 & Given that you have OSS experience, what roles have you played? (Select top 3 roles)* & \MCQ & Shown if $\neq$ ``Never worked'' is not selected for Q10 \\

\addlinespace
Q12 & (Q20) Rate how much each motivation influences your decision to contribute to OSS projects (Not at all--Extremely).* & \LIK & 7-item Likert matrix (e.g., Learning or gaining new skills, enjoyment) \\

\addlinespace
Q13 & How old do you prefer projects to be when you join them?* & \SCQ & -- \\
Q14 & What stage should a project be in when you join?* & \SCQ & -- \\
Q15 & How many contributors should a project have for you to be interested in joining?* & \SCQ & -- \\
Q16 & How many followers/forks (GitHub) should a project have for you to consider joining it?* & \SCQ & -- \\
Q17 & What documentation features do you look for before deciding to join? (Not at all--Extremely).* & \LIK & 4-item Likert matrix (e.g., Clear contribution guidelines provided, The project has its own webpage outside the documentation ) \\

\addlinespace
Q18 & How can project recommendation systems be improved to better align with your motivations? & \FT & Open-ended response \\

\bottomrule
\end{tabularx}
\end{table*}

\end{document}